# MAUVE: An Ultraviolet Astrophysics Probe Mission Concept

Mayura Balakrishnan[1], Rory Bowens[1], Fernando Cruz Aguirre[2,3], Kaeli Hughes[4,5], Rahul Jayaraman[6], Emily Kuhn[7], Emma Louden[8], Dana R. Louie[9,10,11], Keith McBride[5], Casey McGrath[11,12,13], Jacob Payne[3], Tyler Presser[14], Joshua S. Reding[15], Emily Rickman[16], Rachel Scrandis[17], Teresa Symons[18,19], Lindsey Wiser[20], Keith Jahoda[21], Tiffany Kataria[7], Alfred Nash[7], and Team X[7]

[1] Department of Astronomy, The University of Michigan, Ann Arbor, MI, USA
[2] Laboratory for Atmospheric and Space Physics, University of Colorado, 600 UCB, Boulder, CO 80309, USA
[3] Department of Physics and Astronomy, University of Iowa, Iowa City, IA 52242, USA
[4] Department of Physics, Department of Astronomy, Institute for Gravitation and the Cosmos, Pennsylvania State University, State College PA, USA
[5] Center for Cosmology and AstroParticle Physics, The Ohio State University, Columbus, OH 43210, USA
[6] MIT Department of Physics and Kavli Institute for Astrophysics and Space Research, Cambridge, MA 02139, USA
[7] Jet Propulsion Laboratory, California Institute of Technology, 4800 Oak Grove Drive, Pasadena, CA 91109, USA
[8] Department of Astronomy, Yale University, 52 Hillhouse Avenue, New Haven, CT 06511, USA
[9] Catholic University of America, Department of Physics, Washington, DC, 20064, USA
[10] Exoplanets and Stellar Astrophysics Laboratory, NASA Goddard Space Flight Center, Greenbelt, MD 20771, USA
[11] Center for Research and Exploration in Space Science and Technology II, NASA/GSFC, Greenbelt, MD 20771, USA; casey.d.mcgrath@nasa.gov
[12] Center for Space Sciences and Technology, University of Maryland, Baltimore County, Baltimore, MD 21250, USA
[13] Gravitational Astrophysics Lab, NASA Goddard Space Flight Center, Greenbelt, MD 20771, USA
[14] Department of Astronautical Engineering, University of Southern California, Los Angeles, CA 90089, USA
[15] Department of Physics and Astronomy, University of North Carolina at Chapel Hill, Chapel Hill, NC 27599, USA
[16] European Space Agency (ESA), ESA Office, Space Telescope Science Institute, Baltimore, MD 21218, USA
[17] Kavli Institute for Cosmological Physics, University of Chicago, Chicago, IL 60637, USA
[18] Department of Physics and Astronomy, University of California, Irvine, CA 92697, USA
[19] Caltech/IPAC, MC 100-22, 1200 E. California Blvd., Pasadena, CA 91125, USA
[20] School of Earth and Space Exploration, Arizona State University, Tempe, AZ 85287, USA
[21] X-Ray Astrophysics Lab, NASA Goddard Space Flight Center, Greenbelt, MD 20771, USA
Received 2024 June 3; accepted 2024 September 6; published 2024 October 25

## Abstract

We present the mission concept "Mission to Analyze the UltraViolet universE" (MAUVE), a wide-field spectrometer and imager conceived during the inaugural NASA Astrophysics Mission Design School. MAUVE responds to the 2023 Announcement of Opportunity for Probe-class missions, with a budget cap of $1 billion, and would hypothetically launch in 2031. However, the formulation of MAUVE was an educational exercise and the mission is not being developed further. The Principle Investigator-led science of MAUVE aligns with the priorities outlined in the 2020 Astrophysics Decadal Survey, enabling new characterizations of exoplanet atmospheres, the early-time light curves of some of the universe's most explosive transients, and the poorly-understood extragalactic background light. Because the Principle Investigator science occupies 30% of the observing time available during the mission's 5 yr lifespan, we provide an observing plan that would allow for 70% of the observing time to be used for General Observer programs, with community-solicited proposals. The onboard detector (THISTLE) claims significant heritage from the Space Telescope Imaging Spectrograph on Hubble, but extends its wavelength range down to the extreme UV. We note that MAUVE would be the first satellite in decades with the ability to access this regime of the electromagnetic spectrum. MAUVE has a field of view of $900'' \times 900''$, a photometric sensitivity extending to $m_{\rm UV} \leqslant 24$, and a resolving power of $R \sim 1000$. This paper provides full science and mission traceability matrices for this concept, and also outlines cost and scheduling timelines aimed at enabling a within-budget mission and an on-time launch.

*Unified Astronomy Thesaurus concepts:* Ultraviolet astronomy (2170); Space telescopes (1547)

## 1. Introduction

Observations of astrophysical phenomena in the ultraviolet (UV) regime of the electromagnetic spectrum allow us to study our universe on various temporal and spatial scales. However, obtaining UV data can only be done from high altitudes or space due to the Earth's atmosphere's efficient absorption of nearly all ultraviolet photons.

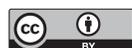

Throughout the past five decades, ultraviolet satellites have revolutionized our understanding of the universe. The first ultraviolet satellite, the Orbiting Astronomical Observatory 2, launched in the late 1960s (Code et al. 1970), enabled early characterization of interstellar extinction (Bless & Savage 1972). Subsequent missions, such as the Extreme Ultraviolet Explorer (EUVE) and the Far Ultraviolet Spectroscopic Explorer (FUSE), were able to study shorter wavelengths. EUVE conducted the first all-sky survey of far-ultraviolet sources and made significant strides in the identification of hot white dwarfs (Vennes et al. 1996), while FUSE—among other discoveries—allowed for some of the first systematic investigations of the nature of the intergalactic medium (Oegerle et al. 2000). The Galaxy Evolution Explorer (GALEX) (Martin et al. 2005) conducted a deeper all-sky survey for UV sources than EUVE did, but did not have detectors capable of reaching into the far- and extreme-ultraviolet regime (it operated between 135 and 300 nm).

Since GALEX was decommissioned in 2013, there have been four primary ultraviolet detectors that NASA commissioned that remain in orbit: the Space Telescope Imaging Spectrograph (STIS), the Wide Field Camera 3, and the Cosmic Origins Spectrograph, all on the Hubble Space Telescope (HST); and the Ultraviolet and Optical Telescope on the Neil Gehrels Swift Observatory (Roming et al. 2005). However, none of these detectors can study the far- and extreme-ultraviolet sky at the precision, spectral resolution, and/or for the duration that is possible with a PI-led mission, necessitating the development of a satellite that can access these regions of the electromagnetic spectrum. We also note that the Habitable World Observatory (HWO) is expected to have UV observing capability. However, HWO is in the early stages of formulation and is not expected to launch until the 2040s.

In this paper, we present the scientific justification for and engineering design of a UV space telescope concept, which responds to the 2023 Draft "Announcement of Opportunity"[22] (AO) for a Probe-class mission with a cost cap of $1 billion (National Aeronautics and Space Administration, Science Mission Directorate 2022). This mission concept was developed as part of the inaugural class of NASA's "Astrophysics Mission Design School" (AMDS),[23] hosted by the Jet Propulsion Laboratory (JPL) in 2023 April. We note that an AMDS ground rule required that our mission not be X-ray or Far IR, bands specifically sought by the AO; eight concepts in these bands were under development for submission in 2023 November concurrent with the AMDS.

Our proposed satellite—"Mission to Analyze the UltraViolet universE," or MAUVE—was an educational exercise developed with the help of Team-X (an engineering team at JPL that specializes in mission concept development). Our proposed mission would be able to access the extreme-, far-, and near-ultraviolet portions of the electromagnetic spectrum to study the transient sky, exoplanets, and the circumgalactic medium. In addition to "Principal Investigator" (PI)-led science cases, as dictated by the AO, 70% of the observing time on this telescope would be allotted to the community via a competitive General Observer program to enable the broadest possible scientific return. The hypothetical MAUVE would launch in 2031. We also note that this mission concept is in no way affiliated with, or derivative of, the Mauve mission (https://bssl.space/mauve/), whose main goal is to perform UV spectroscopy of stars.

MAUVE would fill the gap left in the electromagnetic spectrum at far- and extreme-ultraviolet wavelengths as the first observatory in more than two decades with the ability to detect some of the most energetic ultraviolet photons. Figure 1 shows MAUVE juxtaposed with other NASA-sponsored ultraviolet space missions launched over the past few decades. In this paper, we demonstrate how MAUVE will be able to generate novel scientific results that will, in conjunction with ground- and space-based observatories across the electromagnetic spectrum, unlock a new window on our universe and some of its most unique phenomena. This paper serves as a record of the concept and a resource for those developing future missions who may wish to draw on elements of MAUVE's design and capabilities.

The manuscript is organized as follows. Section 2, Science Investigation, details the PI-led science cases for the mission, while Section 3, Science Implementation, describes the onboard detector technology. Section 4, Mission Implementation, enumerates and defines all the subsystems on board the satellite and describes the Flight Dynamics. Finally, Section 5, Programmatics, details the mission schedule, risks, and anticipated budget/costs.

## 2. Science Investigation

MAUVE's science mission targets five hypothesis-driven objectives spanning the three 2020 Astrophysics Decadal Themes (National Academies of Sciences, Engineering, and Medicine 2021). Figure 2 shows the landscape of these objectives in the context of these themes, the UV spectrum, and the type of target being observed

- *O1.* Determine if sub-Neptune atmospheric escape is caused by photoevaporation or core-powered mass loss.
- *O2.* Determine if the atmosphere composition of hot gas giant exoplanets is influenced by equilibrium or disequilibrium condensation.
- *O3.* Determine if blue kilonovae associated with binary neutron star mergers are driven by radioactive cooling or rapid shock cooling.

---
[22] https://explorers.larc.nasa.gov/2023APROBE/
[23] https://www.jpl.nasa.gov/edu/intern/apply/nasa-science-mission-design-schools/

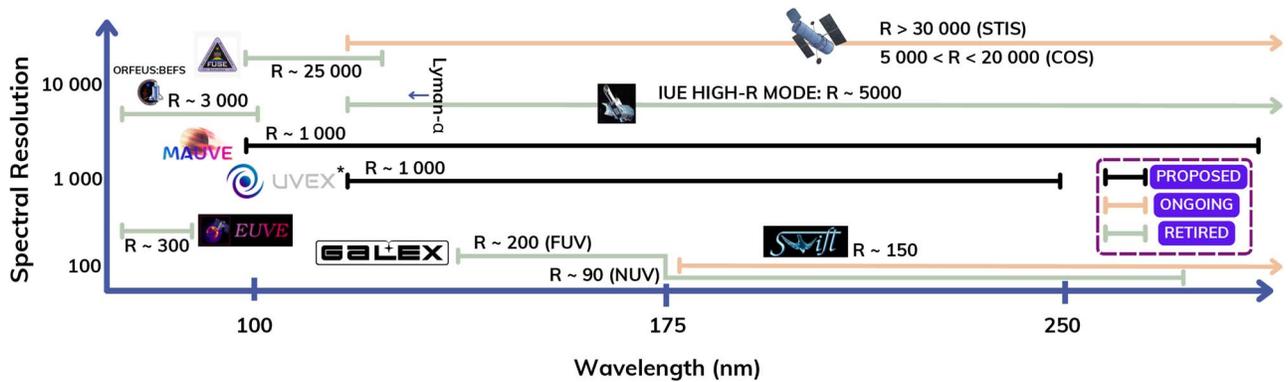

**Figure 1.** An "undiscovered country plot" comparing our proposed mission MAUVE (center) to prior, ongoing, and future NASA-funded missions—including the recently-selected UltraViolet EXplorer (UVEX, Kulkarni et al. 2021). At the time of mission concept design, UVEX was still a proposed mission; since then, it has been selected, so it has been marked with a * to indicate its special status in the landscape of UV missions. Our mission can access a larger wavelength range than previous missions and probe the extreme UV. A space-based detector for extreme-UV photons has not been flown in over two decades.

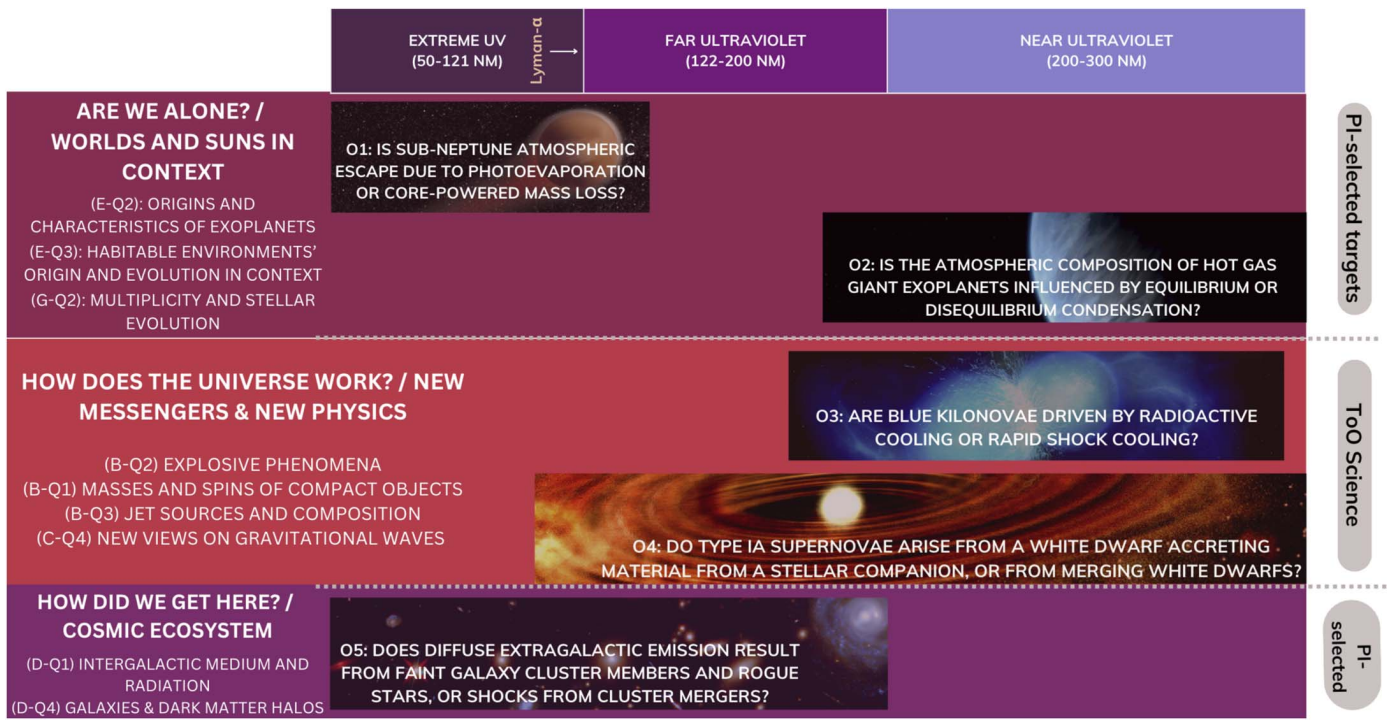

**Figure 2.** The PI-led science landscape of MAUVE. Each objective directly supports priority questions identified from the three themes in the 2020 Astrophysics Decadal. Together, they require a mission that would span the entire UV spectrum from extreme to near-UV. O1 and O2 would observe exoplanets from a pre-selected list of targets, and O5 would observe selected regions on the sky. However, O3 and O4 would follow up on observations of "targets of opportunity" (ToO) as triggers from other observatories are sent.

- *O4*. Determine whether Type Ia supernovae arise from a white dwarf accreting material from a stellar companion, or from the merger of two white dwarfs.
- *O5*. Determine if diffuse extragalactic emission results from faint galaxy cluster members and rogue stars, or shocks from cluster mergers.

Objectives O1 and O2 deepen our understanding of atmospheric composition and evolution of large exoplanets around hot stars. Missions like Kepler (Borucki et al. 2010) and TESS (Ricker et al. 2015) have facilitated the discovery of thousands of exoplanets. MAUVE would allow for the characterization of exoplanets and their evolution by studying atmospheric evaporation, cloud formation, and chemistry. Atmospheric evaporation is one mechanism via which exoplanets could evolve from one class into another during their lifetimes (e.g., from a sub-Neptune into a super-Earth); observing this phenomenon would allow better characterization of the diverse methods through which planets can form. Studying clouds and their chemistry on exoplanets would—just like on Earth—allow us to characterize these planets' climate and weather patterns and investigate aerosol formation and condensation.

Objectives O3 and O4 improve our understanding of some of the most violent explosions in the universe, kilonovae and Type Ia supernovae. Kilonovae (O3) arise from the merger of two neutron stars and are responsible for the production of some of the heaviest elements in the universe. We have a limited grasp on the very early evolution of such transients. Therefore, MAUVE's rapid "Target of Opportunity" (ToO) triggering on kilonovae would allow us to study the cooling mechanisms at play in these events. Type Ia SNe—caused by the detonation of a white dwarf—remain a mystery, as we do not know the true nature of their progenitors. MAUVE's rapid follow-up of these events would allow us to study the interaction of the ejecta with the companion, shedding light on the progenitor systems.

Finally, objective O5 finds the cause of the excess diffuse extragalactic background light (EBL) observed across the UV spectrum throughout the universe. In addition to the light emitted by identifiable galaxies and clusters, there is an excess of light across the EM spectrum from gamma-rays to radio that has been well-characterized by missions and detectors, including Fermi, Swift, HESS, COBE, and ARCADE. However, there exists a notable gap in the UV spectrum for such observations; MAUVE would provide measurements to help fill this gap and, in doing so, further our understanding of stellar and galactic formation across the universe.

Notably, the MAUVE science objectives require complete coverage of the UV spectrum, with spectroscopic/imaging, rapid command/slewing, and fine guidance capabilities. MAUVE's full Science Traceability Matrix (STM) is given in Table 1, which shows how the hypothesis-driven science objectives trace to physical parameters/observables, instrument requirements, and mission requirements. For example, spectroscopic resolution requirements are driven by objectives O1 and O2. Objectives O3 and O4 drive the rapid response capabilities for observing ToO triggers, which require special consideration in the mission observation profile (Section 3.2) and several of the engineering subsystems (e.g., attitude control, flight software—see Section 4.2). Pointing/stability considerations across objectives motivate the inclusion of MAUVE's Fine Guidance Sensor (FGS) (to enable the required pointing stability for long-duration observations).

### 2.1. O1: Determine if Sub-Neptune Atmospheric Escape is Caused by Photoevaporation or Core-powered Mass Loss

Planets with sizes ranging between Earth and Neptune ("sub-Neptunes") are ubiquitous in exoplanet surveys conducted in the last few decades (Bean et al. 2021) despite their absence in our solar system. Understanding how these unique types of planets form is critical to understanding planet formation pathways and sub-Neptunes' potential to support life. Studies of sub-Neptunes with Kepler (Borucki et al. 2010) revealed that their distribution in radii is bimodal (resulting in the Fulton Gap, Fulton et al. 2017) with targets on the smaller side of the distribution labeled "super-Earths" (1–1.75 $R_\oplus$).

One possible hypothesis for this distribution is that all planets on the smaller side of the Fulton Gap are rocky and all originally had gaseous atmospheres while planets on the larger side (radii larger than 2.0 $R_\oplus$) had minimal atmospheric loss (Bean et al. 2021). They thus retained their atmosphere, while planets below the Fulton gap lost their atmospheres due to photoevaporation or core-powered mass loss. Both atmospheric loss processes rely on supplying energy to the planet's atmosphere (via ionizing stellar photons for photoevaporation or cooling luminosity from the planet's core for core-powered mass loss) to allow gas from deeper in the planet's potential well to reach the Bondi radius and escape into the vacuum (Ginzburg et al. 2016). Super-Earths forming from stripped sub-Neptunes may be a key step in producing habitable secondary atmospheres (Luger et al. 2015). This would be relevant specifically to M-dwarf hosts whose habitable zones would overlap with the regions where atmospheric loss due to photoevaporation or core-powered processes would be relevant (Luger et al. 2015; Richey-Yowell et al. 2023).

To understand mass loss, we need to determine the primary mechanism: photoevaporation or core-powered loss. These two processes are differentiable due to different ejection velocities of the gas from their host (Owen et al. 2023). The planet's outflow beyond its Hill radius would take the shape of a cylindrical tail (see Figure 3 for a visual representation). The ejection velocities are small compared to the planet's orbital velocity, making it possible to approximate the outflow as a narrow cylinder along the same semimajor axis.

The central portion of the tail would move away from the planet in the opposite tangential direction of the orbit with the ejection velocity. It is initially optically thick to Ly$\alpha$ and other radiation but optically thin to the ionizing radiation shortwards of 92 nm. The ejection velocity into the assumed cylinder can be solved from the obscuration profile and a known EUV flux (Owen et al. 2023). Core-powered mass loss should result in ejection velocities of around 1–2 km s$^{-1}$ (e.g., Ginzburg et al. 2018) while photoevaporation should result in mass losses closer to 10 km s$^{-1}$ (e.g., Owen & Jackson 2012).

**Table 1**
The Science Traceability Matrix (STM)

| Decadal Goals (Science Panel Questions) | Science Objectives | Science Measurement Requirements — Physical Parameters | Science Measurement Requirements — Observables | Instrument Requirements | Mission Requirements |
|---|---|---|---|---|---|
| (E-Q3) How do habitable environments arise and evolve within the context of their planetary systems? (E-Q2) What are the properties of individual planets, and which processes lead to planetary diversity? (G-Q2) How does multiplicity affect the way a star lives and dies? | O1 Determine if sub-Neptune atmospheric escape is caused by photoevaporation or core-powered mass loss. | **Ejection Velocity** for escaping atmospheres. Velocities for each hypothesis case: 1. *Photoevaporation*: 10-20 km s$^{-1}$ ejection speed. 2. *Core-powered mass loss*: 1-2 km s$^{-1}$ ejection speed. Requires 0-20 km s$^{-1}$ range with 2 km s$^{-1}$ resolution. | **Atmospheric Transmission Spectra** visible for Lyman-α line caused by escaping neutral hydrogen intercepting UV flux from the host star, and transit depths up to 50% absorption with 1% resolution. | Spectrograph with wavelength range 120-123 nm (around Lyman alpha line) with resolution $R = 1000$. Spatial resolution of 7″. | 35 targets chosen from confirmed exoplanets to ensure precise planet parameters. Transits and post-transits cover 4 - 8 hours with ~3 transits per target. Pointing of 0.5″ for stable measurements of the host star. Host stars must be spatially resolved from neighbors. Observations may interrupted and repeated at a later time if needed. |
| | | | **Stellar Flux** that ionizes the escaping planetary atmosphere. Range 50-91.2 nm ± 1 nm (hydrogen dominated atmospheres ionize in this range). | Photometry to resolve host stars with Gaia $m < 15$. Spectrograph with wavelength range 50-92 nm with resolution $R = 42$. | |
| | O2 Determine if the atmosphere composition of hot gas giant exoplanets is influenced by equilibrium or disequilibrium condensation. | **Upper atmosphere molecular composition** along the limbs of gas giant exoplanets with equilibrium temperatures (Teq) 1200-2700 K to support a hypothesis: 1. *Equilibrium condensation*: SiO and Mg gas features at limb temperatures above condensation temperatures (1600 K, but Teq may be cooler). 2. *Disequilibrium/rainout*, trapping molecules in *nightside clouds*: diminished or no SiO and Mg gas features. | **Molecular features in transmission spectra** of gas-phase precursors to prominent Mg$_x$SiO$_x$ clouds: SiO (200-350 nm) and Mg (280-300 nm, multiple peaks) Overlapping features resolved (multiple Fe peaks, 230-280 nm). Resolve the Mg II doublet at 280 nm to account for ISM attenuation. This is the narrowest feature and requires $R = 800$. | Spectrograph with range 200-300 nm and $R = 800$. Spatial resolution 3″ to resolve individul stars from nearest neighbors. Photometry to resolve host stars with Gaia $m < 15$. | Repeated target visits to build SNR ≥ 5. 12 targets chosen with predicatable, regular transit intervals <1 week. Transits and post-transits cover 4 - 7 hours with 4 minute exposures. Estimated 78 days of total observation time. Planets surveyed across Teq's with predicted nightside-only Mg$_x$SiO$_x$ clouds. |
| (B-Q2) What powers the diversity of explosive phenomena across the electromagnetic spectrum? (B-Q3) Why do some compact objects eject material at nearly light-speed jets, and what comprises that material? (B-Q1) What are the mass and spin distributions of neutron stars and stellar-mass black holes? (C-Q4) How will measurements of gravitational waves reshape our cosmological view? | O3 Determine if blue kilonovae associated with binary neutron star mergers are driven by radioactive cooling or rapid shock cooling. | **Energy Release Rate Rise Time** for kilonova cooling mechanisms: 1. *Radioactive cooling*: timescale ~1 day ± 5 hours. Peak energy release rate between 10$^9$-10$^{10}$ ± 10$^9$ erg s$^{-1}$ g$^{-1}$. 2. *Shock Cooling*: timescale 1 hour ± 1 hour. Peak energy release rate between 10$^{10}$-10$^{14}$ ± 10$^9$ erg s$^{-1}$ g$^{-1}$. | **Luminosity** measured in uvm2-band (180-280 nm), and the luminosity range within -17 and -3 AB mag and a resolution of 0.2 mag. 1. *Radioactive cooling*: $m < 24$, with the peak rising above a falling power law spectrum by ~2 mag. 2. *Shock cooling*: $m < 22$ with a smooth power law decay spectrum. | Imager with wavelength range 180-280 nm. Spatial range of 15′ with resolution 1″ pixel$^{-1}$ (resolve point sources in galaxies). Photometric absolute magnitude between -17 and -3 (apparent mag<24) at resolution <0.2 mag. | 25 BNS mergers observations. 6 obs day$^{-1}$ for 3 days, 5 minute exposures. ToO response within 12 hours of trigger. Rapid slew rate (>0.25 deg minute$^{-1}$) to hit 180 degree slew in <12 hours. |
| | O4 Determine whether Type Ia supernovae arise from a white dwarf accreting material from a stellar companion, or from the merger of two white dwarfs. | **Shock Heating** of surrounding material or stellar companion resulting from a collision with SNe Ia ejecta. 1. *Gradually Accreting White Dwarf*: heating timescale ~hours. 2. *Rapid Merger of Binary White Dwarfs*: heating timescale ~days. Requires a temporal range of 0 - 3 days with a resolution of hours. | **Spectroscopic Light Curves** in the 90 - 300 nm bandpass across the first three days, revealing which species drive shock heating excess. 1. *Accreting White Dwarf*: will contain H, He lines throughout the UV. 2. *Binary White Dwarfs*: will not contain H or He lines. Fe and Ni features must be resolved with $R \geq 500$ in the 200 - 300 nm bandpass to distinguish core collapse vs. Ia SNe. | Spectrograph with wavelength range 90-300 nm and resolution $R \geq 500$ at 200 nm. Spatial resolution of 1.28″ to resolve SNe Ia. Photometric range of $m_{uv} \leq 24$. | ≥ 200 SNe observations (≥ 50 Ia total). 2 obs/day for 3 days after slewing to target, ≥ 1 hour exposures (6 hours total per SN). |
| (D-Q1) How did the intergalactic medium and the first sources of radiation evolve from cosmic dawn through the epoch of reionization? (D-Q4) How do the histories of galaxies and their dark matter halos shape their observable properties? | O5 Determine if diffuse extragalactic emission results from faint galaxy cluster members and rogue stars or shocks from cluster mergers. | **Radiance of galaxy clusters** in a sample with radiance between 350 - 400 ± 0.1 photons s$^{-1}$ cm$^{-2}$ sr$^{-1}$ Å$^{-1}$. Comparison of radiance radial profile morphology to established mass-to-light ratios (M/L) and gas density for each cluster will distinguish between hypotheses: 1. *Faint cluster members*: radiance/gas density profiles will track with M/L profile. 2. *Merger shocks*: radiance/gas density profiles will diverge from M/L. | High galactic latitude survey of galaxy clusters with spatial resolution of < 2′ to properly sample the radiance radial profile out to 10 Mpc. | Imager with < 2′ point spread function. Accuracy of 0.1 photons s$^{-1}$ cm$^{-2}$ sr$^{-1}$. Photometric depth to $m_{AB} = 21 \pm 0.1$ (to mask point sources, a confounding factor). Wavelength of < 200 nm (to avoid zodiacal light contamination, a confounding factor). | 150 galaxy clusters observations taken from a high galactic latitude sample ($b > 60°$) 10-30 min. exposures to achieve 350 - 400 photons s$^{-1}$ cm$^{-2}$ sr$^{-1}$ Å$^{-1}$. |

**Note.** MAUVE's five hypothesis-driven science objectives support the indicated 2020 Astrophysics Decadal Science goals and their corresponding science measurement requirements drive the instrument and mission requirements. The right-most column of the STM flows into the left-most column of the Mission Traceability Matrix (MTM), presented later in Table 3.

A mission to determine the ejection velocities relies on two observables: the atmospheric transmission spectra as a function of time and the expected EUV ionization rate based on the host star EUV flux. The atmospheric transmission spectra as a function of time can be observed primarily by monitoring Lyα (121.6 nm) absorption and other key lines (dos Santos et al. 2019; Sing et al. 2019) as in Figure 4.

Host star EUV flux from 50–92 nm is the second essential observable as this regime dominates the hydrogen ionization. An accurate model of hydrogen ionization is necessary to understand the anticipated optical depth with time for the escaping of neutral hydrogen in the planet's tail. Together, we can use the absorption profile with time and the anticipated ionization rate to determine a target's ejection velocity and then overall trends that may indicate the primary cause of atmospheric loss for a collection of targets. The resolution, resolving power, observing time, and other factors are set by the need to accurately determine both observables and potential confounding variables like the interstellar medium (ISM) absorption. By limiting measurements to targets closer than 50 pc, the contamination by the ISM can be mitigated, which could cause misidentification of an escaping atmosphere (Lopez et al. 2019). However, heavy ISM absorption in the 50–92 nm range may preclude a significant number of targets even within a 50 pc radius (see Section 4 of the ESCAPE mission overview France et al. 2021). In a full study, careful pre-selection of viable targets and/or alternative approaches to obtaining 50–92 nm data (such as reconstruction based on even shorter far-UV data) would be necessary. Current missions cannot contribute both the necessary observing time for a large-scale survey and access to the EUV

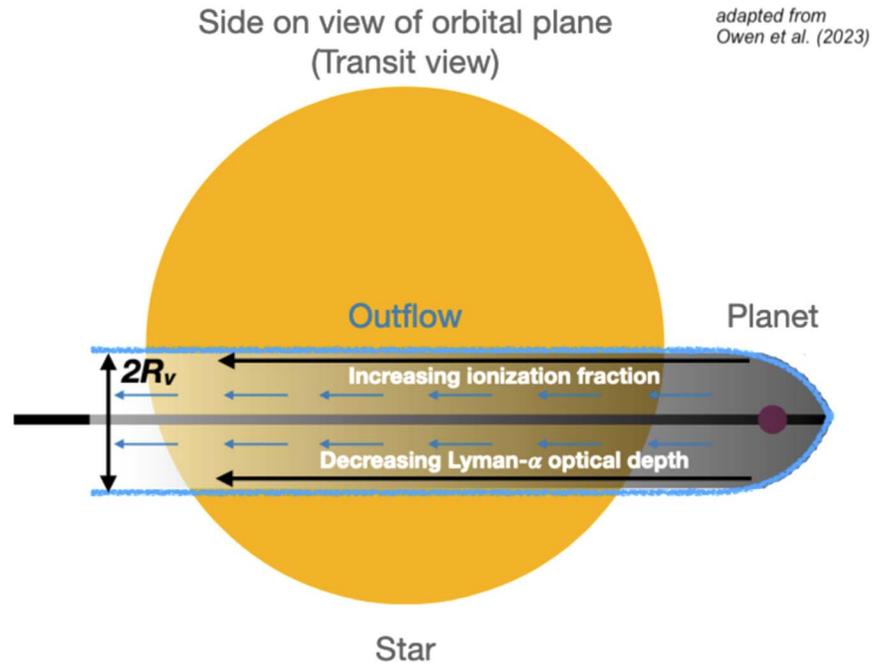

**Figure 3.** A side view of a transit for a planet with an escaping atmosphere. The escaping atmosphere is primarily composed of neutral hydrogen, which is injected into the cylinder at some mass loss rate (a function of several intertwining processes) and with some velocity (the ejection velocity). This neutral hydrogen would move down the tail with its ejection velocity, increasing in ionization fraction as a function of distance and stellar EUV flux. We can measure these observables to find our physical parameter, the ejection velocity, and use that to determine what drives sub-Neptune atmospheric escape. Adapted with permission from [Owen et al. 2023]. © 2022 The Author(s) Published by Oxford University Press on behalf of Royal Astronomical Society.

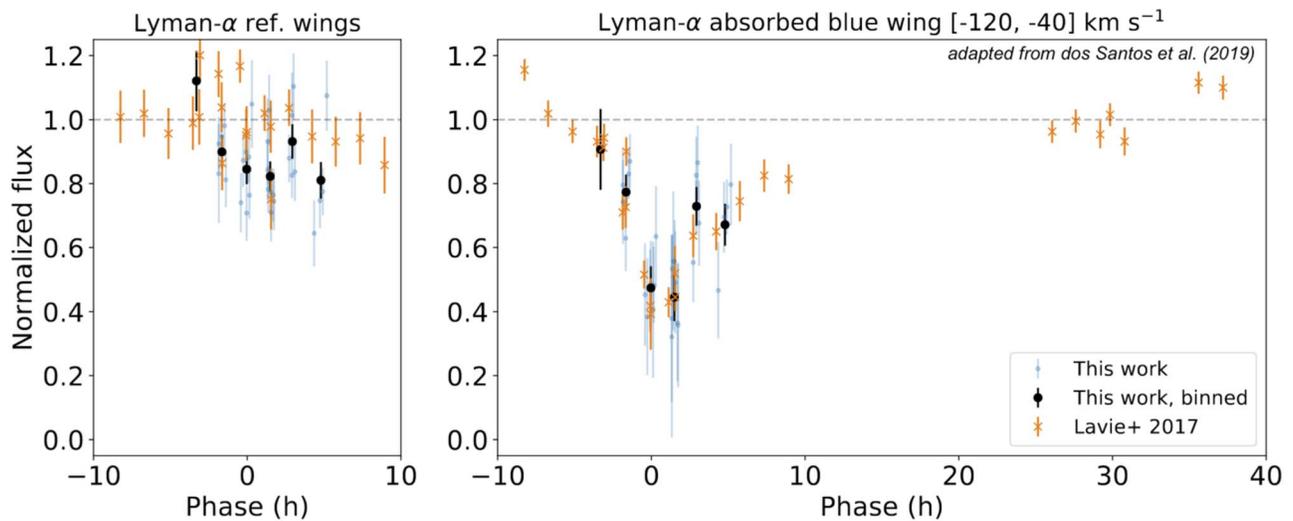

**Figure 4.** Data on the warm Neptune GJ 436b. This data provides important references for one of the observables, the atmospheric transmission spectra. The plot shows normalized flux throughout the transit with 0 h representing when GJ 436b is centrally located on host star. Over the next hour, it leaves the transit, but obscuration remains high due to the escaped neutral hydrogen. As the orbit continues, the hydrogen ionizes, and obscuration quickly declines to 0. Adapted with permission from [dos Santos et al. 2019]. Credit: dos Santos et al., A&A, 629, A74, 2019, © ESO.

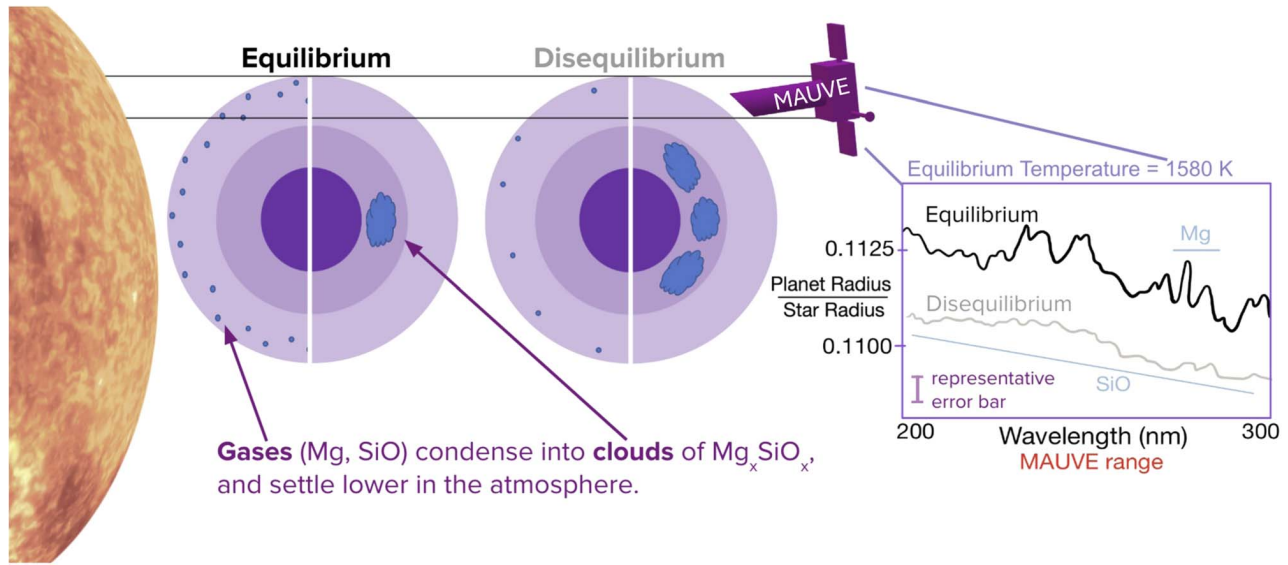

**Figure 5.** A conceptual depiction of the equilibrium and disequilibrium condensation hypotheses. In equilibrium condensation, only local gases become trapped in clouds. In disequilibrium condensation, circulation leads to atoms becoming trapped in clouds and depleting the abundance of their gas states globally. Therefore, the presence or lack thereof of gas-phase precursors in the upper atmosphere indicates equilibrium or disequilibrium condensation, respectively. Example transmission spectra are adapted from models presented in Lothringer et al. (2020). A representative error bar is approximated from simulated observations.

wavelengths needed to accomplish this objective, necessitating the need for a dedicated mission.

### 2.2. O2: Determine if the Atmosphere Composition of Hot Gas Giant Exoplanets is Influenced by Equilibrium or Disequilibrium Condensation

Aerosols (clouds and hazes) are ubiquitous in exoplanet atmospheres (Iyer et al. 2016; Sing et al. 2016), and they significantly impact the chemistry and climate of a planet (Helling 2019). Through spectral observations of hot (>1000 K) gas giants, MAUVE would build our understanding of the processes driving aerosol formation in planetary atmospheres, including thermochemical reactions, photochemistry, and circulation (Marley et al. 2013; Mandt et al. 2022).

From equilibrium chemistry models, estimated atomic abundances, and microphysics models, researchers predict which aerosols exist in a planet's atmosphere at a given temperature and pressure (Gordon & Mcbride 1994; Line et al. 2013; Gao et al. 2020; Goyal et al. 2020). It is hypothesized that atmospheres are relatively clear above 2000 K, silicate clouds are prominent from 1000–2000 K, and hydrocarbon photochemical hazes dominate below 1000 K. We investigate two hypothesized scenarios for aerosol formation in hot atmospheres (see Figure 5):

1. Equilibrium chemistry drives aerosol formation globally. In this case, we expect gas-phase precursors to be distributed throughout the atmosphere at temperatures hotter than the condensation temperature of their respective condensates.
2. 3D variations in temperature and circulation prompt disequilibrium chemistry. In this case, gas-phase precursors would not be abundant at higher temperatures than condensation due to the sequestration of these molecules in locally cooler regions where condensation has occurred.

UV wavelengths probe an atmosphere's low-pressure, high-altitude regions (Wakeford et al. 2018; Christiansen et al. 2019). As aerosols form, we can hypothesize gas-phase aerosol precursors to condense, causing them to settle, or "rain out," of the upper atmosphere, diminishing their spectral features (Lothringer et al. 2020; Gao et al. 2021). This mission proposes to observe the transit spectra of a population of tidally locked hot Jupiters (equilibrium temperatures 1200–2400 K) to obtain positive and null detections of gas-phase condensate precursors—primarily silicate cloud precursors SiO (prominent feature 200–350 nm) and Mg (280–300 nm) (Lothringer et al. 2022). Additional observable features include Fe, Mn, Cr, and Na (Lothringer et al. 2020). These planets have hot daysides and cool nightsides with the potential to sequester molecules, enabling us to trace condensation with planet temperature and differentiate between equilibrium and disequilibrium condensation (Helling 2019; Lothringer et al. 2020; Fortney et al. 2021). The MAUVE mission would provide dedicated time for a large survey in UV wavelengths, including repeated transit observations of planet targets.

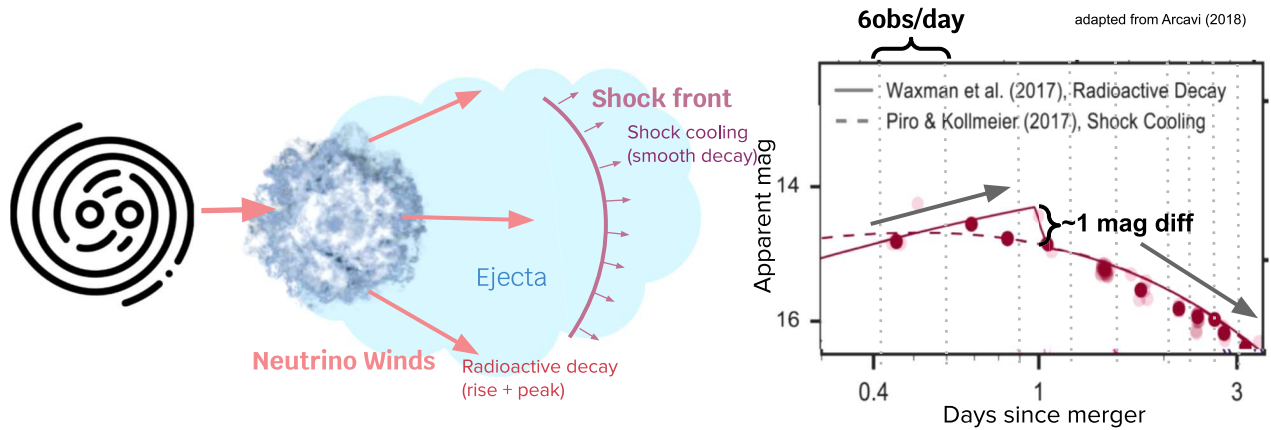

**Figure 6.** Visualization of different cooling mechanisms (left) with UV light curve data (right). The solid line shows the theoretical prediction from a radioactive decay-only cooling mechanism and the shock cooling-only mechanism by the dashed line. Adapted from [Arcavi 2018]. © 2018. The American Astronomical Society. All rights reserved. +B8:M8.

### 2.3. O3: Determine if Blue Kilonovae Associated with Binary Neutron Star Mergers are Driven by Radioactive Cooling or Rapid Shock Cooling

Binary neutron star (BNS) merger events have long been theorized to be environments for some of the most high-energy events in space (Metzger 2017). On 2017 August 17th, LIGO and VIRGO recorded their first gravitational wave (GW) signal from a BNS merger (Abbott et al. 2017). An extensive electromagnetic (EM) follow-up campaign followed the GW alert, and the merger results found an early peak in the UV a few hours after the merger event (Barnes 2020). The source of this peak is unknown; to be "blue," i.e., to peak in UV, the material emitting the light must be composed of lighter elements (Aprahamian et al. 2018). Furthermore, since the blue kilonova comes after the merger, the matter emitting the light must be cooling. This indicates two possible main mechanisms for the emission of a blue kilonova:

1. *Radioactive Decay.* After the event, if the resulting merger product is a massive neutron star, it will emit strong neutrino "winds" (Waxman et al. 2018). Some of this wind will interact with the surrounding ejected mass, inducing radioactive cooling. The cooling will emit strongly in the near-UV (180–280 nm) and peak ∼2 mag above a power law cooling spectrum roughly a day after the merger.
2. *Shock Cooling.* Post-merger, a cocoon of ejected mass will surround the merger product. Given the energetics surrounding the merger, an associated γ-ray burst (GRB) is expected, triggered by the collapse of a neutron star into a black hole (Piro & Kollmeier 2018). This GRB will produce a shock front that heats the surrounding ejecta and emits strongly in the near-UV. The cooling ejecta will be brightest immediately after the shock front passes and will dim following a falling power law spectrum within the luminosity range of −17 to −3 AB mag.

Note that although different merger remnants drive the two cooling mechanisms, they give rise to blue kilonova light curves that are very similar ∼1 day after the merger event, as shown in Figure 6 (Arcavi 2018).

However, shortly after the merger, a rapid UV follow-up could indicate which cooling mechanism dominates the creation of the blue kilonova. Breaking this degeneracy has many scientific implications, including constraining the neutron star equation of state (EOS) by probing the merger progenitors' masses (Margalit & Metzger 2017). Constraining the EOS has wide-reaching implications in nuclear and condensed matter physics, therefore representing one of the more prominent questions in astrophysics.

MAUVE would probe this cooling degeneracy by observing LIGO/VIRGO/KAGRA BNS merger triggers multiple times in the UV band shortly after the merger event. The resolution, or lack thereof, of the ∼1 day luminosity peak, indicative of radioactive cooling, would point to the dominant cooling mechanism of the merger.

### 2.4. O4: Determine Whether Type Ia Supernovae Arise from a White Dwarf Accreting Material from a Stellar Companion, or from the Merger of Two White Dwarfs

Type Ia supernovae (SNe Ia) are among the most scientifically valuable events in modern astronomy not only due to their fundamental physics (Höflich 2006), but also because their behavior offers useful context for other fields of inquiry, including stellar evolution (Liu et al. 2022; Swastik et al. 2022; Rajamuthukumar et al. 2023), galactic evolution (Stinson et al. 2006; Dwek 2016), nuclear physics (Höflich 2006; Tumino et al. 2018; Mori et al. 2019), GW studies (Korol et al. 2017; Zou et al. 2020), and especially in cosmology as "standard candles" for measuring cosmic distances (Riess et al.

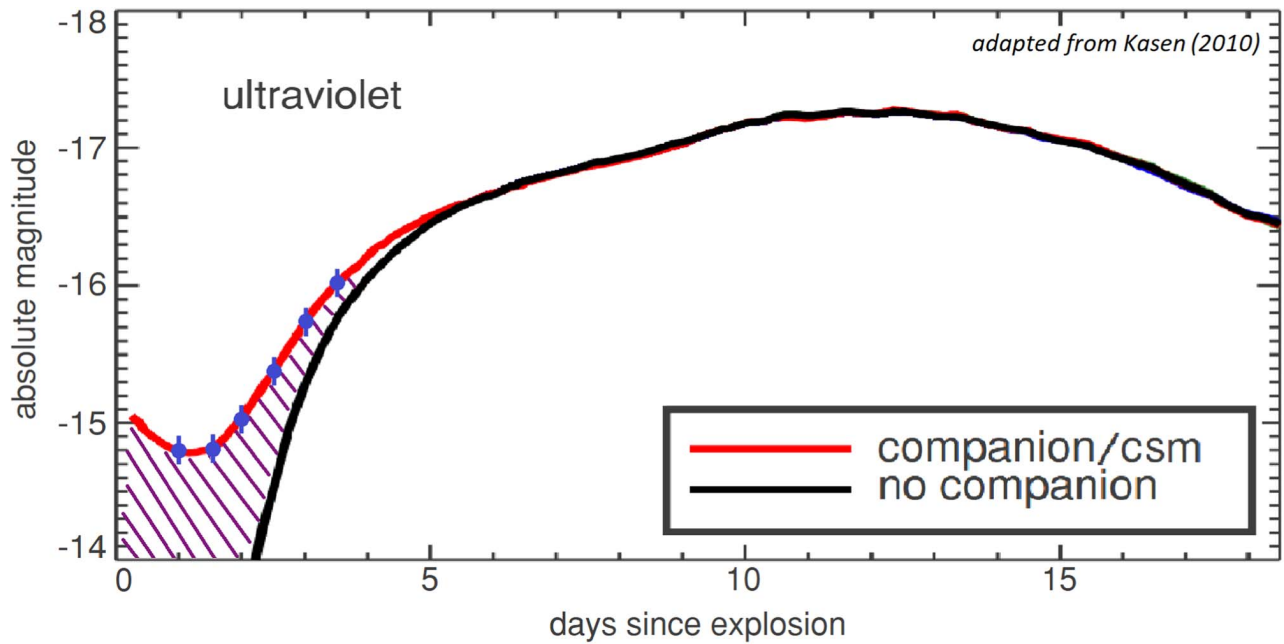

**Figure 7.** Early UV excess from shock heating. Light curves of SNe Ia show that the shock heating signature is strongest in the UV (90–300 nm) within the first few days after detonation. An example of the photometric shock heating signature from a companion or CSM is shown in red, while a theorized curve for no ejecta impact is shown in black. The MAUVE observation cadence is shown in blue; spectroscopic features (i.e., the presence or absence of H, He) observed in this early time frame would indicate an SD or DD progenitor mechanism. Adapted from [Kasen 2010]. © 2010. The American Astronomical Society. All rights reserved.

1998, 2004; Meng et al. 2015; Brout et al. 2019). However, because we cannot predict when a detonation will occur and consequently have not yet observed a progenitor system, its detonation, and its subsequent remnant in any wavelengths, we do not know the dominant progenitor mechanism(s) for SNe Ia. Though we know that these explosions result from the runaway thermonuclear detonation of a carbon-oxygen core white dwarf (WD), the scenarios that instigate these detonations (each of which carries various theorized nuances) may either be:

1. *Single-degenerate (SD).* gradual accretion from a main sequence or giant companion (Whelan & Iben 1973). In this case, we expect to detect strong signatures of H or He from the donor star.
2. *Double-degenerate (DD).* rapid and violent merger of two WDs (Webbink 1984; Maoz & Hallakoun 2017). In this case, MAUVE would marginally detect H or He, as WDs are mostly stripped of these elements.

A plausible model must explain both the observed galactic occurrence rate and the apparent ∼70% uniformity of SNe Ia detonations (Li et al. 2011).

MAUVE would be uniquely equipped to slew to SNe Ia and observe in the UV (90–300 nm) hours to days after detonation when critical spectral signatures would imply either the SD scenario or the DD scenario. As ejecta collides with a companion star or circumstellar medium (CSM) containing the remnants of the donor, a visual magnitude excess may appear and peak within 3–48 hr post-detonation (Kasen 2010). By integrating spectroscopic observations collected during this early light curve phase, MAUVE would both detect whether this brightness excess occurs and characterize the former donor, thereby indicating the likely progenitor scenario.

MAUVE would be capable of making 2 observations per day for the first 3 days after slewing to an SN target, with the trigger informed by ground-based transient alert systems, e.g., the Transient Name Server (TNS). We note that alerts cannot distinguish between Ia and core-collapse SNe in these early evolutionary stages, but SNe can be selected from general alerts by associating locations with known galaxies using, e.g., the NASA/IPAC Extragalactic Database (NED). All SN alerts would be observed with low-resolution ($R = 500$) spectroscopy collected in 1 hr exposures (6 hr total per SN); Figure 7 displays a simulated observation of SN Ia ejecta with MAUVE. Within a volume of 100 Mpc ($z = 0.02$), driven by the instrument's limiting magnitude, we expect 400 SNe to occur each year; of these, ∼75% are expected to be core-collapse SNe, while the remaining 25% would be SNe Ia (Graur et al. 2011; Li et al. 2011). Selecting the highest-quality 10% of candidates required to produce a statistically significant sample, 200 ToO triggers (∼50 SNe Ia total; 1200 hr on-sky) would be observed over the 5 yr nominal lifetime of the mission. Data collected for core-collapse SNe would be made available to the community and would not use observing time reserved for General Observer programs (see Section 2.6).

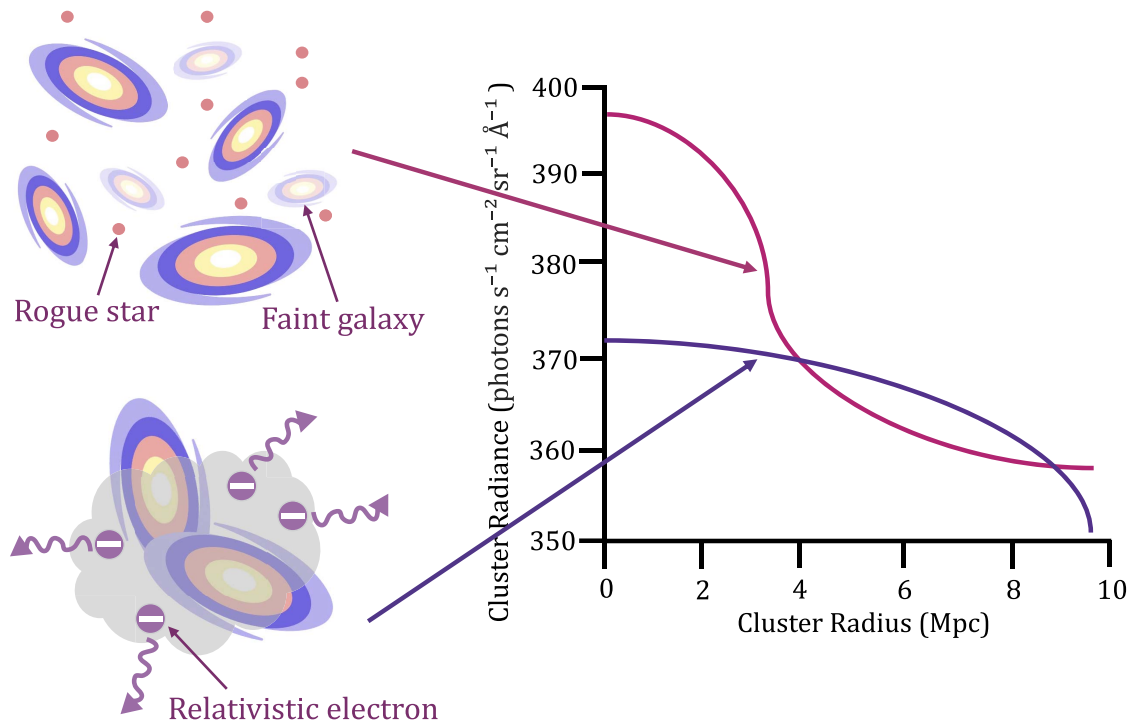

**Figure 8.** A comparison of the radial profiles of cluster radiance with mass and gas density would allow us to discern the source of excess UV emission from galaxy clusters: faint cluster members or relativistic electrons from merger shocks (Welch et al. 2020).

### 2.5. O5: Determine if Diffuse Extragalactic Emission Results from Faint Galaxy Cluster Members and Rogue Stars, or Shocks from Cluster Mergers

Diffuse background radiation resulting from extragalactic sources, referred to as the EBL, exists across the entire electromagnetic spectrum. Measurements of the EBL spectrum provide constraints on models of galaxy formation and evolution (Cooray 2016), a key area of focus in the 2020 Decadal Survey (National Academies of Sciences, Engineering, and Medicine 2021). Due to the absorption of extragalactic photons by both neutral hydrogen in the Milky Way and the intergalactic medium, measurements of the ultraviolet EBL (referred to as the cosmic UV background—CUVB) have been particularly challenging compared to those made at other wavelengths (Cooray 2016; Murthy et al. 2019).

While some components of the CUVB are galactic in origin, a significant component is not able to be explained by galactic processes. This portion of the CUVB is likely produced by extragalactic processes such as active galactic nuclei and star-forming galaxies. A discrepancy between the expected emission from these known sources and CUVB measurements indicates the existence of unknown sources of diffuse emission not currently resolved by telescopes (Cooray 2016). Massive galaxy clusters are one such candidate source of this emission.

The excess in CUVB emission measured in these clusters is possibly sourced from (see Figure 8):

1. *Faint cluster members.* These include both galaxies that are unresolved and intracluster light, also referred to as intrahalo light (Welch et al. 2020). Intrahalo light is the light emitted from ejected or rogue stars between galaxies, which have been difficult to detect due to their faint nature (Cheng et al. 2021).
2. *Shocks from mergers.* Relativistic electrons may be produced in the ISM as a result of shocks from merging galaxy clusters (Welch et al. 2020).

By learning the source of emission within these clusters, we learn if this emission is the product of the earliest star formation in the universe or a natural part of the evolution of galaxies.

Measuring the radial profile of cluster radiance for a sample of galaxy clusters allows for an external comparison to mass and density profiles from existing data sets (Welch et al. 2020). If the source of excess UV emission is faint cluster members, the radiance and gas density profiles would closely match the mass-to-light profile. Alternatively, if the source is relativistic electrons in the gas between clusters resulting from shocks, the radiance and gas density profiles would diverge from the mass-to-light profile due to cluster activity (Clowe et al. 2004).

MAUVE would observe a sample of 150 galaxy clusters out to a cluster radius of 10 Mpc for 30 minutes each in order to gain sufficient resolution in cluster stacking. MAUVE's bandpass avoids contamination from zodiacal light present at $\lambda > 250$ nm (Murthy et al. 2019), while MAUVE's pixel scale

**Table 2**
Possible General Observer Program Investigations

| Science Case | Decadal Priority Questions |
|---|---|
| Studying hot massive stars and the most massive and magnetic white dwarfs. | **(G-Q1)** What are the most extreme stars and stellar populations? |
| | **(G-Q2)** How does multiplicity affect the way stars live and die? |
| Supermassive black hole accretion. | **(B-Q4)** What seeds supermassive black holes, and how do they grow? |
| Core-collapse supernovae. | **(B-Q2)** What powers the diversity of explosive phenomena across the electromagnetic spectrum? |
| Observations of Solar System planets (e.g., aurorae on Jupiter). | **(E-Q2)** What are the properties of individual planets, and which processes lead to planetary diversity? |
| M-dwarf flares and their impacts on planetary systems. | **(E-Q3)** How do habitable environments arise and evolve within the context of their planetary systems? |
| Mechanics that control star accretion and formation. | **(F-Q3)** How does gas flow from parsec scales down to protostars and disks? |
| | **(G-Q2)** How does multiplicity affect how stars live and die? |
| Studying starburst galaxy environments. | **(F-Q1)** How do star-forming structures arise from and interact with the diffuse ISM? |
| Investigating Quasars. | **(B-Q3)** Why do some compact objects eject material at nearly-light-speed jets, and what is that material made of? |

**Note.** Additional example science cases identified as potential areas of investigation of MAUVE's GO program, and the corresponding decadal questions that they would support.

provides the necessary resolution of the inner cluster cores, which have not been properly sampled in previous UV surveys (Welch et al. 2020). Such a measurement would resolve the source of diffuse UV extragalactic emission in galaxy clusters and elucidate cluster formation and evolution. Neither currently operating missions nor available archival data have the required resolution to capture the radial profile of inner cluster cores.

### 2.6. General Observer Program

The design capabilities of this mission (driven by the PI-led science) would enable the "General Observer" (GO) program to investigate many additional key questions from the Decadal Science Goals. This is a key consideration of MAUVE's mission profile, given that the GO program would constitute 70% of the observation time (see Section 3.2).

During its operation, MAUVE would solicit proposals for observation time that would then be scheduled throughout the PI science observations. While the breadth of proposed science objectives would ultimately be determined by the community, we have identified several example science cases in Table 2 that would support additional Decadal Goals beyond the goals already supported by the PI-science (see Table 1).

Additionally, O4 offers a unique opportunity for a cross-over between PI-lead science and GO-enabled science. MAUVE requires observations of SNe Ia transients, but it is expected that these will only constitute ∼25% of the total observations of SNe that it makes, while the other ∼75% are expected to be core-collapse SNe. Unfortunately, MAUVE cannot discriminate in advance of the observation of the supernova type and hence cannot prioritize only observing the SNe Ia transients. However, the additional core-collapse data taken by MAUVE during PI-science observing time would be made public to the wider scientific community. This enables a significant science contribution to the astrophysics community which would not have to be directly solicited via proposals through the GO program.

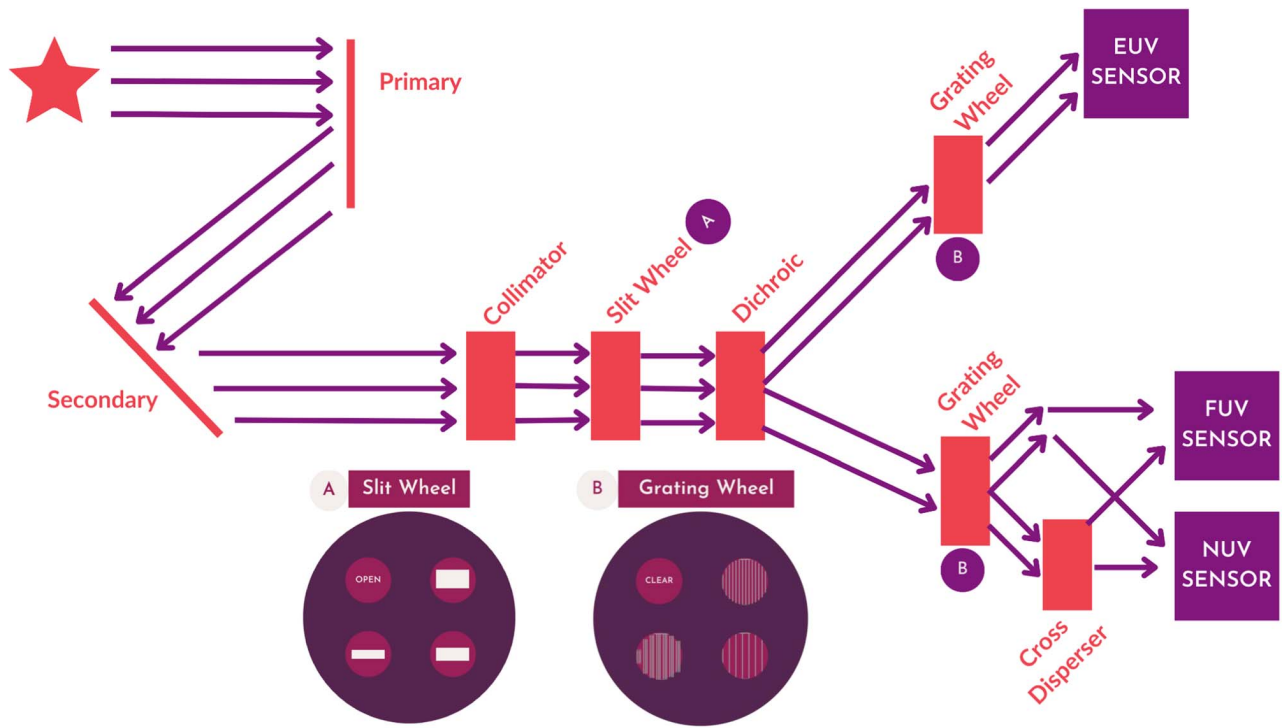

**Figure 9.** The optical path of the THISTLE instrument. Light is first collected by the MAUVE telescope. The light is collimated and then passes through a slit wheel (A). A dichroic is implemented to separate the light into EUV and FUV-NUV channels. A grating wheel (B) is used along both channels to transition between imaging and spectroscopy modes. A cross disperser is available in the FUV-NUV channel for echelle spectroscopy. THISTLE would be operable in 32 configurations.

## 3. Science Implementation

This section presents the capabilities of the imaging/spectroscopy instrument on MAUVE, and presents a science mission profile that will allow MAUVE to achieve its scientific goals, even if certain observations are interrupted by ToO triggers. We also present a data format and archiving plan that facilitates easy, rapid public access to MAUVE data.

### 3.1. Instrumentation

MAUVE's primary instrument is the Hybrid Imager/SpecTrograph for uLtraviolet into Extreme wavelengths (THISTLE), designed to meet the science objectives laid out in Section 2. THISTLE would be capable of imaging and spectroscopy in the EUV (50–121 nm), FUV (122–200 nm) and NUV (200–300 nm) wavelengths. The optical path of THISTLE is summarized in Figure 9. A slit wheel is implemented to enable slitted spectroscopy. A dichroic separates light into EUV and FUV-NUV channels. A grating wheel is present in both channels to alternate between imaging and spectroscopic modes. In the FUV-NUV channel, echelle spectroscopy is enabled by including a cross disperser. Light is ultimately focused onto three sensors, optimized to capture the light in each UV bandpass.

The optical design of THISTLE shares significant heritage with two previously flown UV spectrographs: the STIS[24] on board the HST, and the Berkeley Extreme and Far-UV Spectrometer (BEFS)[25] on board the Orbiting and Retrievable Far and Extreme Ultraviolet Spectrograph (ORFEUS) space shuttle mission. The overall optical design is similar to STIS, observed in FUV, NUV, and optical bandpasses. The optical channel in the STIS design is replaced with the optical design of the BEFS EUV channel.

The inclusion of an EUV channel is driven primarily by O1 to capture ionizing radiation from exoplanet host stars. As the sensitivity of the optical elements could vary be several orders of magnitude across the overall bandpass, careful characterization should be performed to ensure the final performance is viable to achieve the objective. Slitless NUV spectroscopy, required for O2, drives THISTLE's NUV performance. An imaging mode is included in the optical design to conduct the high galactic latitude survey required for O5. MAUVE will use a FGS identical to the FGS on board JWST to achieve the target stability.

---
[24] https://www.stsci.edu/hst/instrumentation/stis
[25] https://archive.stsci.edu/missions-and-data/orfeus/befs

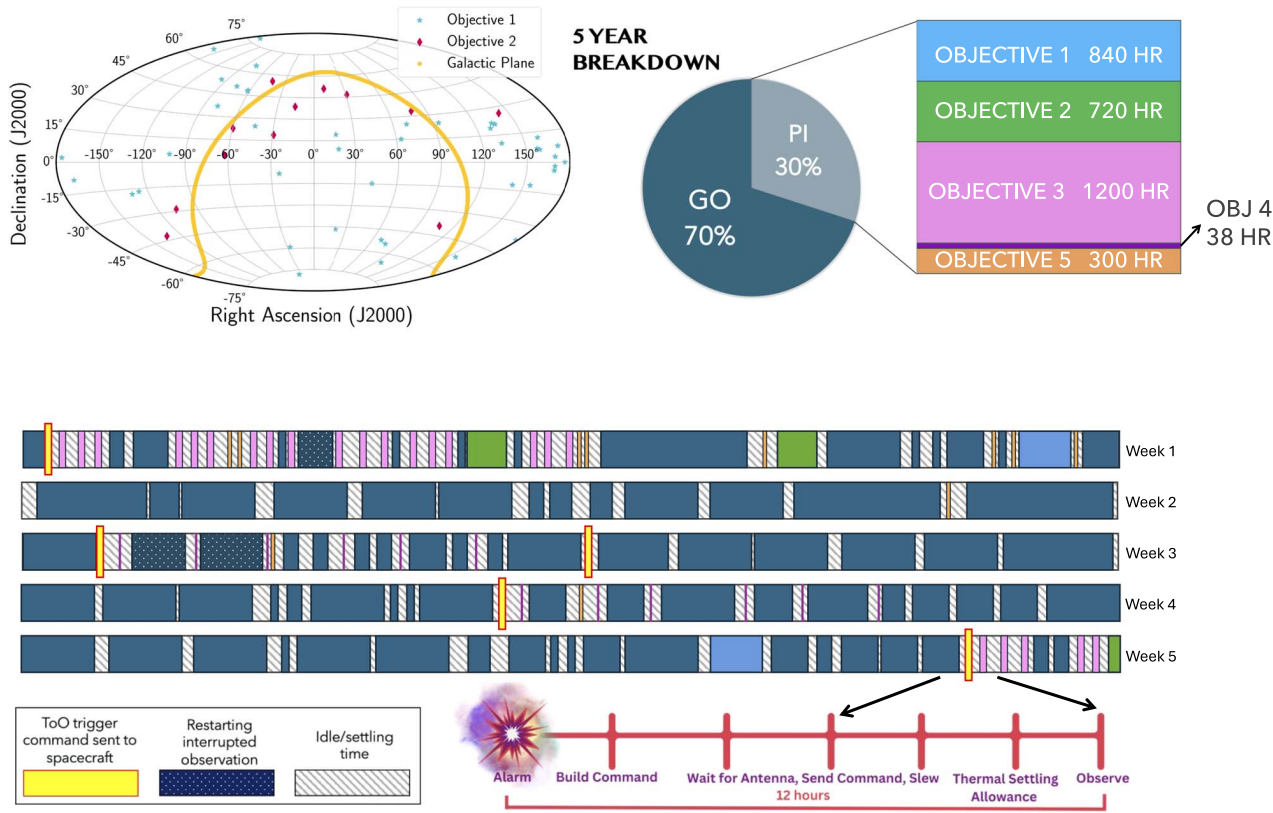

**Figure 10.** TOP LEFT: Sky map showing targets selected to complete Science Objectives O1 and O2. TOP RIGHT: A breakdown of the 30% of PI mission time by objective. We do not show the time allotted for slewing, settling, and instrument idle time (∼6000 hr) or remedying disruptions by ToO observations (∼2790 hr). BOTTOM: Example observation schedule over five weeks. The colors correspond to the colors in the pie chart. We include blocks for ToO triggers, restarted observations, and settling/slewing time. At the bottom right, we show the process that would initiate when a ToO alarm goes off, where the probe would only slew if we can begin observing within 12 hr.

### 3.2. Science Mission Profile

MAUVE is designed to achieve the science closure of all PI-science objectives within five years while allowing 70% of the observation time to fall under a GO program. While O1, O2, and O5 would be achieved with scheduled exposures, O3 and O4 would be completed solely through ToO triggers. The mission will operate at L2 to allow exposures of the entire sky throughout the mission's existence; an Earth orbit would result in eclipses, but operating at L2 allows a larger sky area to be visible constantly. A short required exposure time means that observations for O5 will not be interrupted and can be easily slotted in among others. An algorithm will be necessary to determine whether ToO triggers should be acted upon. With an interval of 8 hr between downlinks, the maximum time between a trigger and the beginning of the observation would be 12 hr. An example observation schedule lasting five weeks is shown in Figure 10, highlighting our capability to maintain a schedule even if half of the O1 and O2 exposures are interrupted by a ToO trigger. Within the 30% allotted time for P.I. science, 3096 hr would be used for observations, 2790 hr for restarting interrupted observations, and 6930 hr for slewing, assuming a conservative slew time of 30 minutes driven by thermal limitations (Section 4.2). We are still able to maintain a margin of ∼300 hr for unexpected issues.

### 3.3. Data Sufficiency, Analysis, and Archive Plan

The STM (Table 1) provides details on the data required for science closure. For a resolution of $R = 1000$ and a frame size of $2048 \times 2048$ pixels, each frame is expected to need $3.28 \times 10^7$ bits of data. PI science would result in 1.6 Tbits of data. To ensure rapid response to transient events, MAUVE will communicate with our ground systems every eight hours. Assuming three downlinks a day, achieving science closure requires an average downlink rate of $0.2$ Mb s$^{-1}$, well within our telecommunication and Ground Data Systems (GDS) capabilities (Section 4.2.3). The MAUVE mission strategy involves hosting lower-level L0–L1 data products (raw science data and

Table 3
The Mission Traceability Matrix (MTM)

| Mission Requirements | | Mission Design Requirements | | | Spacecraft Requirements | | Ground Systems Requirements | | | Operations Requirements | |
|---|---|---|---|---|---|---|---|---|---|---|---|
| STM (O1) | **MR1:** 35 target exoplanets with visually distinct stars. **MR2:** Pointing Control better than 0.5" | NNH22ZDA015J (B-33) | **MDR1:** Compatability with Launch Vehicle | MR2 | **SR1:** Three Axis Stabilized | NNH22ZDA015J (36) | **GSR1:** Downlink: Ka-Band | MR8 | **OR1:** Commanding Frequency: > 1 per 12 hours | | |
| STM (O2) | **MR3:** 12 target exoplanets with transit frequency > 1 week$^{-1}$. **MR4:** Repeated exoplanet target observations, transits/post-transits covering 4-8hrs, build SNR ≥ 5. **MR5:** Planets surveyed across $T_{eq}$'s with predicted nightside-only $Mg_xSiO_x$ clouds. | 2023 Astro Probes LV Summary_Rev3 | **MDR2:** Mass Allocation: ≤ 3376 kg | MDR2 | **SR2:** Carry the THISTLE Instrument & MAUVE Telescope Basic Mass: 631 kg | NNH22ZDA015J (36) | **GSR2:** Uplink: X-Band | GSR3 | **OR2:** Orbit Determination sufficient to maintain Telecommunications and L2 station keeping | | |
| STM (O3) | **MR6:** 25 Binary NS merger observations. **MR7:** 6 obs day$^{-1}$ for 3 days of ToOs, ≥ 1 hour exposures. **MR8:** ToO response within 12 hours of trigger. **MR9:** Slew rate ≥ 0.25deg min$^{-1}$ | NNH22ZDA015J (103) | **MDR3:** Launch Date: NLT January 2032 | MR4 | **SR3:** Instrument accommodation: Fine Guidance Sensor Thermal Stability (<1°C fluctuation across the optical instrument during data acquisition) | NNH22ZDA015J (37) | **GSR3:** Maintain connection to NASA Lunar Exploration Ground Sites (LEGS) with 3 dB link margin | | | | |
| STM (O4) | **MR8 & MR9** also trace from O4 **MR10:** ≥ 200 SNe observations (≥ 50 Ia total) | 2023 Astro Probes LV Summary_Rev3 | **MDR4:** L2 Halo Orbit (Launch C3 ≤ -0.7 km² s$^{-2}$). | MDR6 | **SR4:** Downlink rate: >1092 kbits s$^{-1}$ | | | | | | |
| STM (O5) | **MR11:** 150 galaxy clusters observations taken from a high galatic latitude sample ($b$ > 60°). **MR12:** 10-30 minute exposures to achieve 350 - 400 photons s$^{-1}$ cm$^{-2}$ sr$^{-1}$ Å$^{-1}$. | NNH22ZDA015J (Sections 4.1.4 and 5.1.4) | **MDR5:** Mission Length: ≥ 64 months | MDR1 | **SR5:** 52Ah battery to handle the highest expected depth of discharge at launch | | | | | | |
| | | NNH22ZDA015J (B-21) | **MDR6:** 30% PI mission time (including contingencies/margins), 70% GO mission time. | MR7 | **SR6:** 1582W (solar) to provide 20% margin over EOM average wattage use case | | | | | | |

**Note.** The MTM records the mission functionalities flowing from major requirements to achieve science objectives in the STM (Table 1) within the bounds of program constraints from the Draft AO and Launch Vehicle Services (NASA Launch Vehicle Services Program 2023). The left-most column of the MTM flows from the right-most column of the STM.

calibrated light curves and spectra) on the Mikulski Archive for Space Telescopes (MAST). More advanced L2 products would be available within one month, with all reductions carried out via the CALSTIS pipeline (Katsanis & McGrath 1998).

## 4. Mission Implementation

In this section, we present the mission trajectory of MAUVE, along with details regarding all the subsystems to ensure that all the mission goals are realized over its lifetime. A summary of the mission implementation is given in the MAUVE Mission Traceability Matrix in Table 3.

### 4.1. Mission Design, Trajectory, and Launch Vehicle

MAUVE would arrive in a halo orbit at L2 after a month of travel to orbit insertion (Figure 11). As noted earlier, L2 allows exposures of the entire sky throughout the mission's existence; an Earth orbit would result in eclipses, but operating at L2 allows a larger sky area to be constantly visible. After insertion, it will be 0.012 au from Earth. The fuel budget allows for three trajectory correction maneuvers (TCMs). At decommissioning, a $\Delta V$ of 5 m s$^{-1}$ will move the telescope away from L2 stability.

A single launch aboard any standard-performance launch vehicle described by the Draft AO can accommodate the MAUVE spacecraft. The specifications for the LV flow down from the STM into the second column of Table 3. MAUVE would be inserted at L2 utilizing a LV with $C_3 = -0.7$ km$^2$ s$^{-2}$ that meets the MPV (with propellant) of 1900 kg.

### 4.2. Flight Systems

Our flight systems were designed with the help of software from Team X. MAUVE's systems exceed the stipulated requirements and ensure the mission remains over Technology Readiness Level 6.

#### 4.2.1. Configuration, Thermal, and Power

*Mechanical and Configuration.* A compact rectangular bus contains all instrument and flight system components, including the FGS, which is displaced in Figure 12 for visibility. All components are symmetrically balanced for stability during launch. Two solar panel wings containing two panels per wing are attached to the bus, and each wing actuates on a 1 degree-of-freedom gimbal.

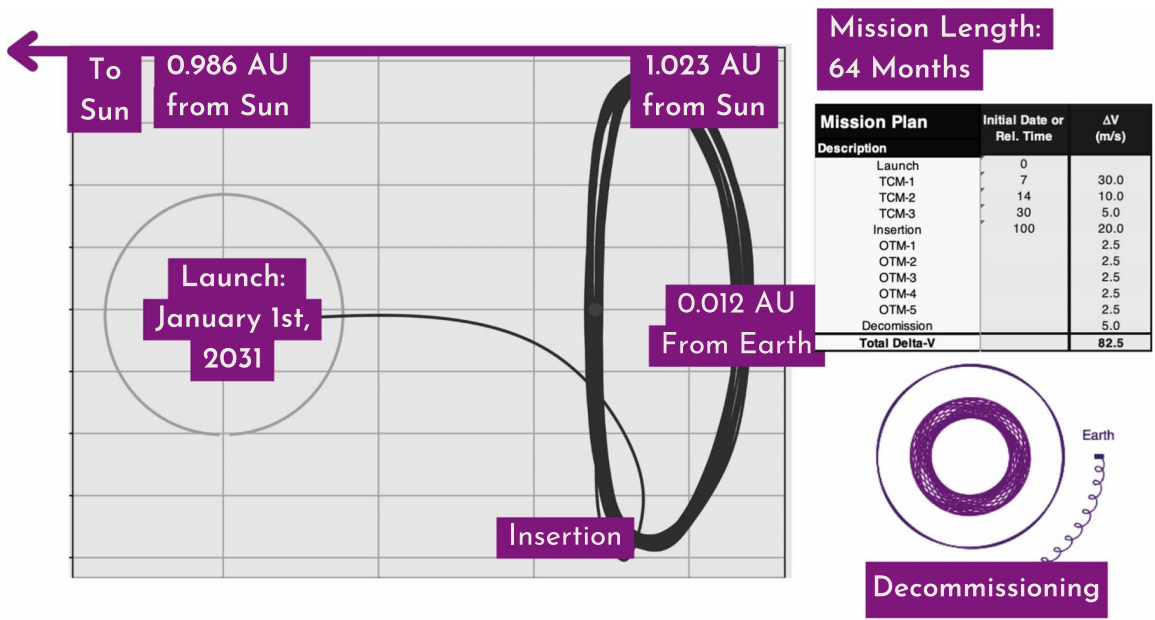

**Figure 11.** MAUVE's launch, insertion, orbit, and decommissioning plan.

*Power.* The dual-stringed solar power system at L2 with 4.17 m$^2$ active area can provide 1582 W at the beginning of life and 1387 W at the end of life, resulting in a 20% margin on the end of mission requirement. This should allow the power system to handle the average needed wattage beyond the 5 yr mission for up to 9.7 yr. A 52 Ah capacity battery at the beginning of the mission is sized to handle the highest expected depth of discharge (60%) at launch.

*Thermal.* The telescope optics are built to withstand temperatures between −20°C and +45°C, and are planned to be held at −15°C during operation. Critically, the telescope optics must also be stable, with less than 1°C of fluctuation across the optical instrument during data acquisition. This can be achieved with a cold-biased thermal system, which will electrically heat the critical components like the propellant and optics while radiating additional heat created by the avionics. A thermal blanket will provide additional insulation and temperature regulation throughout the instrument.

#### 4.2.2. Maneuvering

*Propulsion.* The propulsion system consists of twelve monopropellant thrusters. Four main thrusters are used for the three TCMs, insertion into L2, and decommissioning, as shown in Figure 11. The remaining eight thrusters are used for station keeping at L2 and desaturating the ACS reaction wheels. The current design of the propulsion system would support up to a 10 yr mission lifetime.

*Attitude Control System (ACS).* The MAUVE ACS system meets control requirements with dual redundancy and heritage. Attitude control and stability are maintained with a fully redundant ACS system. The system is comprised of Sun sensors, star trackers, reaction wheels, gyroscopes, and the instrument-integrated FGS. The ACS system is designed for three modes of operation: Safe, Nominal, and Science. Sun sensors handle guidance during Safe Mode to re-establish orientation in the event of an anomaly and by use of the reaction wheels to reorient the spacecraft. Nominal operations can be completed using the star trackers and gyroscopes for guidance.

To achieve its science objectives, MAUVE requires both rapid slewing capabilities (for ToO response) and stability (to stay locked on targets). MAUVE plans to incorporate the FGS used on JWST to meet the fine-pointing accuracy required to conduct science. The FGS directly integrates with the science instrumentation and uses a pick-off mirror from the instrument light path. The ACS system, as designed, would achieve the technical requirement of 0.25 deg minutes$^{-1}$ slew and 0.1″ pointing with an additional margin.

#### 4.2.3. System Control and Communication

*Flight Software.* The flight software (FSW) that will be implemented for MAUVE draws from significant JPL Core FSW heritage. Called flIght softwaRe Interfacing Subsystem (IRIS), many components will be reused from previous flights as seen in Figure 13. Given data volume, no compressing or processing onboard is expected to be necessary. The instrument interfacing will be completely new, as is standard for missions. IRIS will also deviate from JPL's Core FSW in its telecomm interfacing as well. In order to achieve ToO capabilities at the desired low latency, the telecomm software will require the ability to handle priority commands in order to interrupt current measurements for higher priority events.

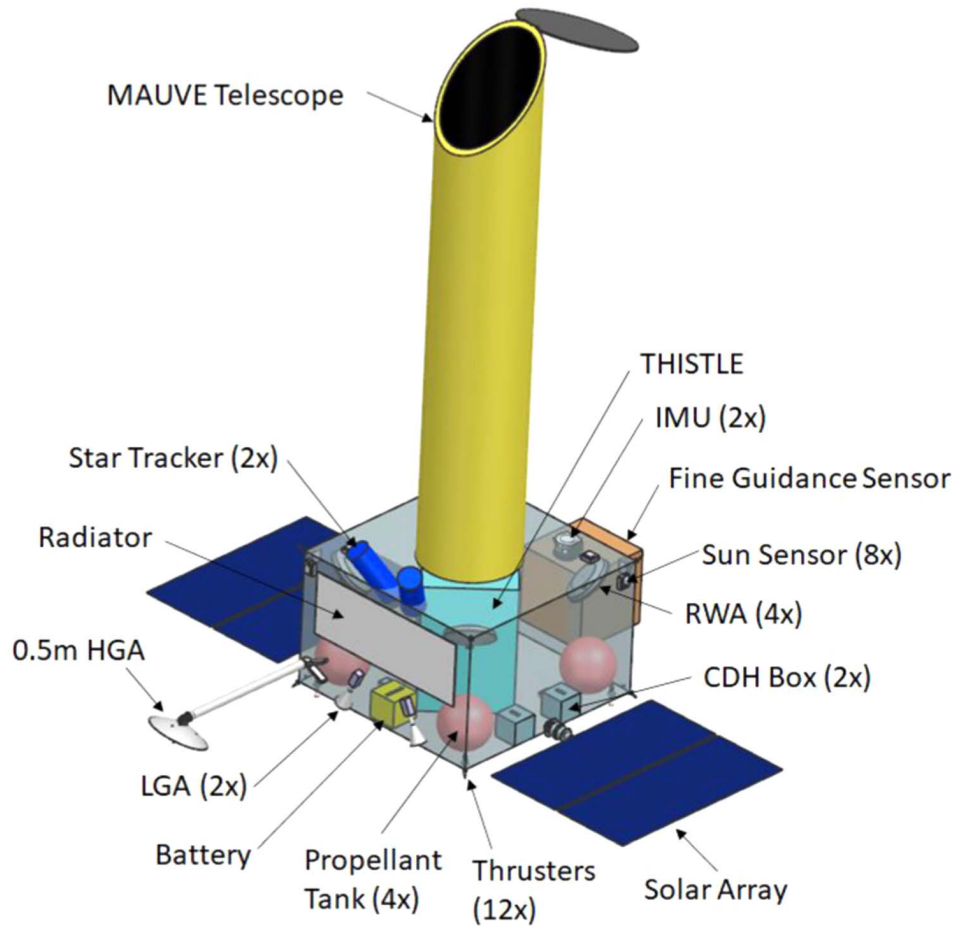

**Figure 12.** Subsystems of the MAUVE spacecraft, pictured with the Fine Guidance Sensor, visually shifted out from the primary instrument and with a translucent chassis for clarity. MAUVE is designed as a rectangular chassis with 3-axis stabilization, single-axis solar array drives, and a prominent, asymmetric telescope baffle. Subsystems include The Hybrid Imager / SpecTrograph for ultraviolet into Extreme wavelengths (THISTLE), Inertial Measurement Units (IMUs), Reaction Wheel Assemblies (RWAs), Command and Data Handling (CDH) Boxes, a Low Gain Antenna (LGA), and a High Gain Antenna (HGA).

*Command and Data Systems.* The minimum specifications for data acquisition are described in Section 3.3. The command and data handling system used in the Mars Science Laboratory (MSL) (Grotzinger et al. 2012) meets these requirements; MSL enables the collection of 300 GB per day of science data, meeting all mission data requirements with a 840% margin. The command and data handling system for this mission utilizes the MSL technology as it meets the requirements and has flight heritage.

*Telecommunications.* A dual-string telecommunications system completes our required downlink in one hour per day with 30% overhead (see Section 3.3 for requirements). We design for ground communication every eight hours to enable rapid response to ToOs. A gimbaled antenna enables simultaneous science observations and communication. *X*-band uplink and *Ka*-band downlink, with secondary *X*-band downlink, are compatible with Lunar Exploration Ground Sites (LEGS) and required link margins.

*Ground Data System.* The MAUVE GDS, illustrated in Figure 14, includes the mission operations center (MOC) that communicates with MAUVE via LEGS and sends data to the science data system (SDS). The science operations center (SOC) directs the MOC, receives ToO alerts, and interacts with GOs and SDS. The SDS likewise interacts with GOs and sends final data products to the MAST at STScI for archival. LEGS enables data downlink in 1 hr per day, allowing for retransmission every 8 hr as needed. The Deep Space Network (DSN) provides a viable backup to LEGS if necessary.

A potential MAUVE mission enhancement proposes that universities build, fund, and operate dishes to communicate with the spacecraft. This would provide an educational opportunity for students while decreasing MAUVE's demand on LEGS and increasing spacecraft commanding opportunities, shortening MAUVE's response to ToOs to less than 8 hr. University dishes would receive encrypted commands directly from the MOC and automatically transmit those commands to the spacecraft with, for example, uplink speeds of 2 kbit s$^{-1}$ for a 9 m dish. Participating universities would gain student enrichment through engineering and astronomy opportunities and potential participation in the MAUVE science collaboration. A functional example of this enhancement is currently in practice at Morehead State University in Kentucky, where a 21 m dish operates as the first non-NASA affiliated node of the DSN (Malphrus et al. 2022).

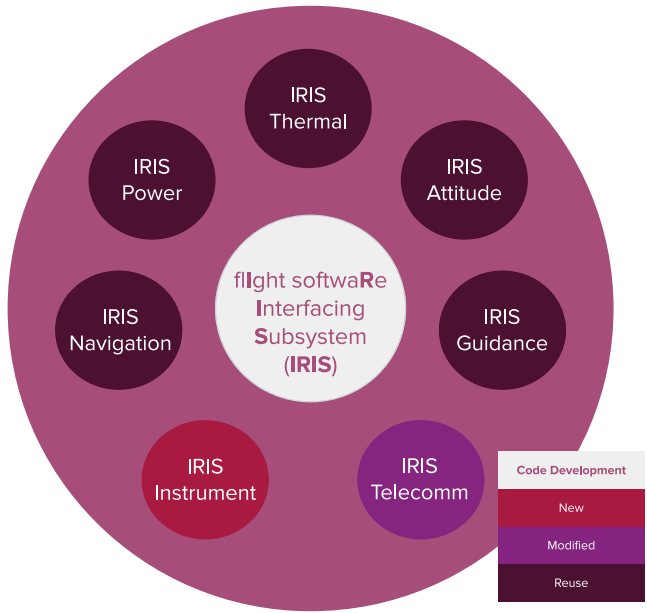

**Figure 13.** Components of IRIS and their development requirements.

## 5. Programmatics

In the following subsections, we describe the notional schedule of our mission design, the Team X validated cost associated with our mission design as well as the perceived mission risks and risk mitigation strategy. All of the following cost information is of a budgetary and planning nature and is intended for informational purposes only.

### 5.1. Schedule

Our schedule, with reserves, allows for a hypothetical targeted launch of 2031 January in accordance with the AO (see Figure 15). Phase A (Concept and Technology Development) is 12 months, as specified in the AO. Phase B (Preliminary Design and Technology Completion) is scheduled for 13 months, allowing 12 months of notional schedule time and 1 month of fully funded schedule reserve. Phase C (Final Design and Fabrication) is 24 months, which includes a 22 months notional schedule plus 2 months of fully funded schedule reserve. Phase D (System Assembly, Integration, and Test) is 22 months, including 4 months of fully funded schedule reserve. Phase E (Science Operations) is 64 months in total (including a 4 months cruise to L2 and 60 months of science operations), which is in accordance with the AO of a minimum of 60 months of science operations. Phase F (Closeout) is 4 months including final archival of data.

All schedule reserve is fully funded and taken into account in the estimated cost (see Section 5.3). This robust schedule is based on a database of historical JPL missions, where the median value of schedule length for each phase is taken. The schedule reserve alleviates the programmatic risks associated with the schedule as outlined in Section 5.2.

### 5.2. Risks and Mitigation Strategy

No major risks were identified during the study. Three moderate risks and two minor risks were identified spread across the implementation, development, and operations stages of the mission. We report the following moderate risks associated with the MAUVE mission architecture in the implementation and development stages:

1. Supply issues could cause delays in the procurement of sensors required for the instrument and telescopic components. To mitigate this, the contract for procurement of sensors and telescopic components would be prepared in Phase A. This will allow the contract to be signed and initiated at the beginning of Phase B, and allow additional time for procurement.
2. Thermal Vacuum Chamber (TVAC) testing delays causing the launch date to be pushed. To mitigate this, additional time has been added to the schedule as fully funded schedule reserves ("Work Breakdown Structure" (WBS) 5, see Section 5.3) to account for any delays related to TVAC.
3. The model for thermal stability needs to be refined to calculate the thermal settling time required for ToO observations. To mitigate this, there will be an assessment of ToOs for follow-up with prioritization of targets within 40° of current observations to reduce the thermal settlement time. Time has been allotted in the schedule in Phase A to run thermal simulations to understand and verify what the thermal settlement time is. Even if the thermal settlement time is at the maximum, the science mission profile outlined in Section 3.2 is feasible and has been calculated to take into account frequent ToO triggers.

We report the following identified minor risks in the science operations category:

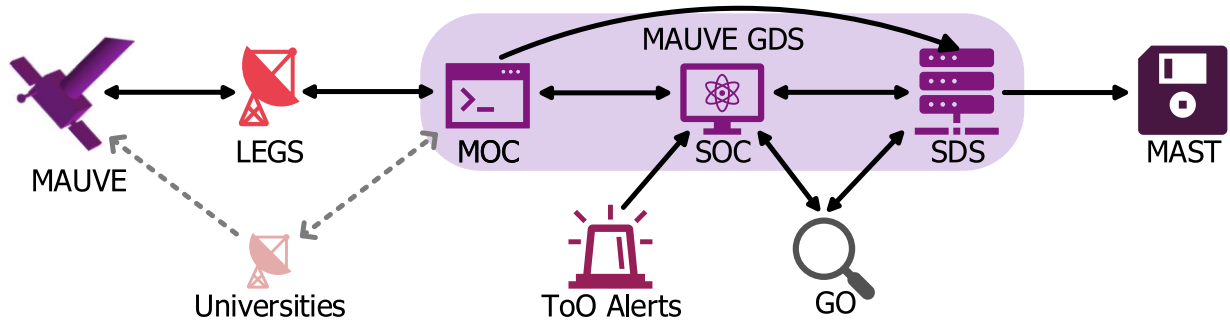

**Figure 14.** The MAUVE Ground Data System (GDS) communicates with the spacecraft via LEGS, receives and acts on ToO alerts, interfaces with GOs, and archives data at MAST. Communication using university-managed dishes to increase ToO response time and offload from LEGS is a possible mission enhancement.

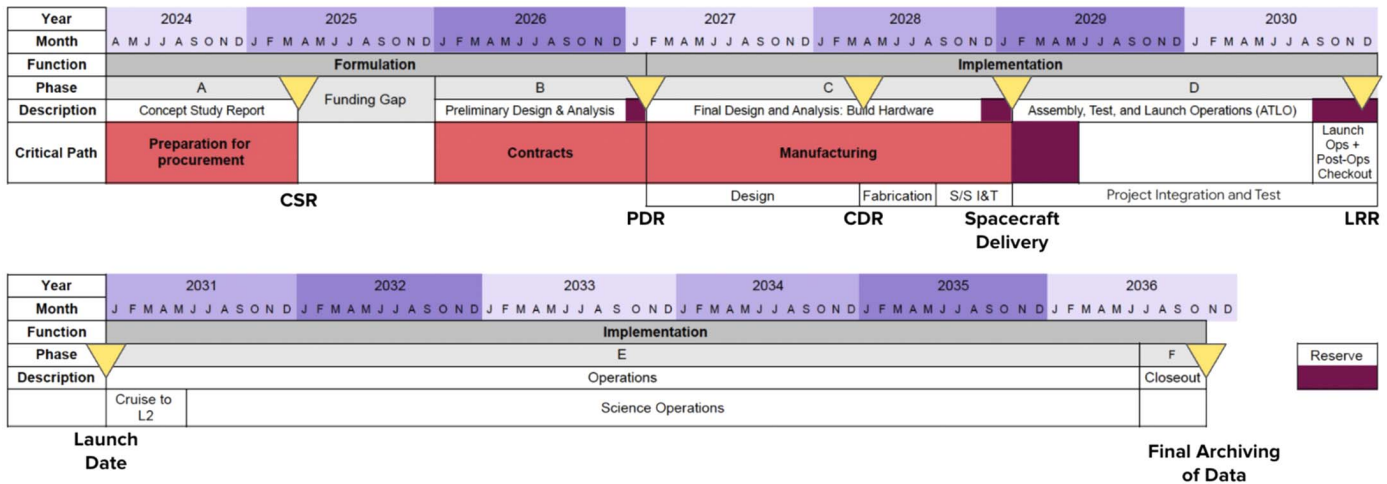

**Figure 15.** Outline of the hypothetical schedule for a targeted launch of 2031 January per the AO. The gray bars highlight the phases of the mission development, the dark purple highlight corresponds to the scheduled funded reserve to account for any delays, the red highlights the critical path of the mission development, and the yellow arrows correspond to the major milestones throughout the mission development schedule. This includes the concept study report (CSR), preliminary design review (PDR), critical design review (CDR), spacecraft delivery, launch readiness review (LRR), the launch date, and the final archiving of the data.

1. Handling disruptions from ToOs for science operations. To mitigate this, additional support will be added to handle disruptions if current science operations are not set up to accommodate low latency response to ToO. This is accounted for in our cost risks under WBS4 (see Section 5.3), with a potential upper of approximately $2.5–5M for Full-Time Equivalents (FTEs).

2. Planning and sequencing of ToOs that are not yet accommodated in the current configuration operation for PI-led and GO Programs. To mitigate this, additional support will be added for the planning and sequencing of ToOs that are not yet accommodated in the current configuration operation for PI-led and GO programs. This is accounted for in our cost risks under WBS7 and WBS9 (see Section 5.3), with a potential upper of approximately $2–3M for mission systems, FTEs, and software.

These plans allow us to properly mitigate all of the associated programmatic risks identified with the MAUVE mission architecture. With the mitigation strategy in place, all of the programmatic risks are minor and have a criticality of "monitor." Where necessary, the mitigation strategies that involve cost are outlined in Section 5.3.

*5.3. Cost*

Our estimate for the PI-managed mission cost (PIMMC, Table 4) accounts for our development strategy, as well as requirements stipulated in the AO. Our mission's estimated cost of $979M is within the $1B Cost Cap specified in the AO. For Developmental phases B–D, we provide generous cost reserves of 30%, exceeding the 25% AO requirement. Our cost reserves for Operational Phases E and F are 10% for Science and 25% (per AO) for all other "Work Breakdown Structure" (WBS) elements. We estimated our costs based upon out-of-house Spacecraft Development, designing to meet the requirements for Mission Class C (per the draft AO), but adding redundancies. Phase A is capped at $5M per the AO.

Table 4
Top Level Summary of MAUVE PI-Managed Mission Costs ($M FY23)

|  | Current Best Estimate | Reserves | Projected Best Estimate |
|---|---|---|---|
| **Total Phases A-F** | **763** | **216** | **979** |
| *Phase A* | 5 | 0 | 5 |
| Total Phases B - D (Development) | 656 | 196 | 852 |
| *Phase B* | 66 | 30% | 85 |
| *Phases C/D* | 590 | 30% | 85 |
| *Launch Vehicle* | 0 | ... | 0 |
| Total Phases E/F (Operations) | 102 | 20 | 122 |

Table 5
MAUVE's Work Breakdown Structure (WBS) Estimated Costs, Bases of Estimates, and Cost Validation

| WBS Number | Purpose | Phases B-D (Development, $M FY23) | Phases E/F (Operations, $M FY23) | Bases of Estimates | Cost Validation |
|---|---|---|---|---|---|
| 1.0 | Project Management | 30 | 6 | JPL Institutional Cost Model (ICM)<br>• Historic missions of similar class/size<br>• In-house development at JPL | Historic actuals such as Spitzer and Kepler |
| 2.0 | Project Systems Engineering | 24 | 0.2 | | |
| 3.0 | Mission Assurance | 26 | 0.8 | | |
| 4.0 | Science | 18 | 30 | Historic missions. Estimates driven by:<br>• Number of instrument sensors<br>• Science teams<br>• Observations for PI-led science (30% of mission)<br>• General Observer science (70% remainder) paid for by NASA | Consistent against actuals within the same mission class for Astrophysics Missions |
| 5.0 | Payload System | 220 | 0 | • THISTLE: NASA Instrument Cost Model (NICM)<br>• MAUVE Telescope: Parametric, Stahl's Telescope Model<br>• Fine Guidance Sensor: JWST actuals | Modeled in NICM, validated against STIS telescope + similar instruments w/ similar UV detectors |
| 6.0 | Flight System | 246 | 0 | JPL ICM: Model-based quasi-grassroots, estimated by subsystem Subject Matter Expert based on historic actuals | Consistent and in line with the expected cost for the flight system budget of a $1B class mission |
| 7.0 | Mission Operations | 27 | 55 | Historic in-house astrophysics observatory missions, including operations, planning, and sequence | Consistent and validated against previous GO program expectations (such as those for Spitzer) |
| 9.0 | Ground Data Systems | 19 | 10 | | |
| 10.0 | Assembly, Test, Launch, Operations | 34 | 0 | JPL ICM: historic missions, in-house development, as in WBS 1-3 | Consistent, and in line with the expected cost for the flight system budget of a $1B class mission |
| 12.0 | Mission and Navigation Design | 10 | 0 | JPL ICM: historic missions; inputs to destination, planning, and navigation | Consistent with previous historic missions |

Table 5 shows our estimated costs by WBS element, alongside the corresponding bases of estimate and cost validations. We incur no costs associated with the launch vehicle (WBS 8.0), since our launch requirements fall well within the NASA-provided standard launch services. Additionally, WBS 11.0, Education and Public Outreach, is not a part of PIMMC.

Our cost estimates are based on historic missions and institutional cost models (ICM), to include the NASA ICM (Habib-Agahi et al. 2011) and the JPL ICM employed by Team X. We rely upon parametric estimates and the Stahl Model (Stahl 2019) for the MAUVE telescope assembly, and we use JWST actuals for the FGS. The JPL ICM algorithms have been developed by the responsible line organizations based on actual mission costs for their area of applicability, such as ACS, CDS, GDS, etc. The models have been validated against a set of mission actuals that were not used in the development of the models themselves. They have a prediction error in aggregate of about 30% for total mission cost, which is comparable to other early formulation cost models. The JPL ICMs undergo updates and independent review approximately every 5 yr, and they last underwent this process within a year of our mission design.

## 6. Conclusion

Photometry and spectroscopy in the UV regime are powerful methods for probing and understanding important astrophysical objects, structures, and transient events. Currently, HST hosts three of the four currently operating UV instruments; the Operational Paradigm Change Review, underway at the time of writing, aims to prioritize HST's unique UV capabilities. However, as HST continues to age, limitations to observations, such as the recent transition to Reduced Gyro Mode, make the future of UV observations with HST uncertain. Therefore, it is important to prioritize the development of smaller-scale missions to fill the UV gap between HST and the next large observatory expected to have high-sensitivity, broadband UV capabilities, the Habitable Worlds Observatory (The LUVOIR Team 2019; Gaudi et al. 2020). Upcoming missions, including ULTRASAT and UVEX, will aid in filling the UV gap between HST and MAUVE provides another mission concept to fill this important gap.

The MAUVE mission concept, while not currently planned to be further developed, provides an example of what a dedicated Probe-class platform for conducting UV astrophysics might look like. MAUVE aims to analyze the UV universe, from exoplanets to explosions, across a broad bandpass (50–300 nm) with both imaging and spectroscopic modes. MAUVE's science program addresses all three themes outlined in the 2020 Astrophysics Decadal (National Academies of Sciences, Engineering, and Medicine 2021) and would provide answers on the driving mechanisms of sub-Neptune atmospheric escape (O1), cloud formation of hot gas giants (O2), blue kilonovae (O3), SNe Ia (O4), and diffuse extragalactic emission (O5). In addition, 70% of MAUVE's observing time would be reserved for GO programs, making MAUVE a formidable community asset for conducting UV investigations. A variety of GO science cases, such as studying supermassive black holes, solar system bodies, stellar UV activity, starburst galaxies, and much more, would be made possible with a platform like MAUVE. A unique science implementation plan was developed for MAUVE; ToOs would be allowed to interrupt scheduled observations, with a margin built into the schedule to resume interrupted observations. Although there is not currently an AO for which MAUVE may be proposed, we present this mission concept to serve as a source of inspiration for UV astrophysics missions of all sizes.

## 7. Disclosures

The cost information contained in this document is of a budgetary and planning nature and is intended for informational purposes only. It does not constitute a commitment on the part of JPL and/or Caltech. The information provided is pre-decisional and for planning and discussion purposes only.

## 8. Materials Availability

Materials for this work were prepared during the AMDS at JPL with proprietary tools and the help of Team X. Therefore not everything is available. Upon reasonable request, inquiries to the availability of data presented in this paper may be made to the authors.

## Acknowledgments

This research was carried out in part at the Jet Propulsion Laboratory, California Institute of Technology, under a contract with the National Aeronautics and Space Administration (80NM0018D0004). We acknowledge AMDS Administrator Joyce Armijo, AMDS School Manager Leslie Lowes, JPL's Team X, and those who reviewed this proposal. We thank the NASA HQ Science Mission Directorate and the NASA Astrophysics Science Division for providing financial support for the inauguration of this AMDS in JPL's Mission Design School portfolio.

The authors would like to recognize and thank the contributions of Dr. Emily Gilbert, who also participated in this Astrophysics Mission Design School, and in particular offered excellent insight into the MAUVE instrument designs.

C.M. and D.R.L. would like to acknowledge support by NASA under award number 80GSFC21M0002. D.R.L. also acknowledges research support by an appointment to the NASA Postdoctoral Program at the NASA Goddard Space Flight center (GSFC), administered by Oak Ridge Associated Universities (ORAU) under contract with NASA.

Finally, Figures 6 and 8 contain graphical elements from Flaticon.com. Figure 1 contains mission icons/logos.

## ORCID iDs

Mayura Balakrishnan ⓘ https://orcid.org/0000-0001-9641-6550
Rory Bowens ⓘ https://orcid.org/0000-0003-0949-7212
Fernando Cruz Aguirre ⓘ https://orcid.org/0000-0003-4628-8524
Kaeli Hughes ⓘ https://orcid.org/0000-0002-4551-9581
Rahul Jayaraman ⓘ https://orcid.org/0000-0002-7778-3117
Emma Louden ⓘ https://orcid.org/0000-0003-3179-5320
Dana R. Louie ⓘ https://orcid.org/0000-0002-2457-272X
Keith McBride ⓘ https://orcid.org/0000-0003-2195-4324
Casey McGrath ⓘ https://orcid.org/0000-0002-6155-3501
Joshua S. Reding ⓘ https://orcid.org/0000-0003-1862-2951
Emily Rickman ⓘ https://orcid.org/0000-0003-4203-9715
Teresa Symons ⓘ https://orcid.org/0000-0002-9554-1082
Lindsey Wiser ⓘ https://orcid.org/0000-0002-3295-1279
Keith Jahoda ⓘ https://orcid.org/0000-0003-0100-6415
Tiffany Kataria ⓘ https://orcid.org/0000-0003-3759-9080